\documentclass[11pt,a4paper]{article}

\usepackage[a4paper,margin=1in]{geometry}
\usepackage{amsmath}
\numberwithin{equation}{section}
\usepackage{amssymb}
\usepackage{graphicx}
\usepackage{authblk}
\usepackage{hyperref}
\usepackage{relsize}
\usepackage{mathtools, nccmath}
\usepackage{xcolor}
\usepackage{cancel}
\usepackage{float}
\usepackage{array, booktabs, makecell}
\usepackage{siunitx}
\usepackage{physics}
\usepackage{placeins}
\usepackage{tikz}
\usepackage{rotating} 
\usepackage{xcolor}
\usepackage{array}
\usepackage{tabularx}
\usepackage{makecell}
\usepackage{adjustbox}

\usetikzlibrary{arrows.meta,positioning,calc,fit,backgrounds}
\graphicspath{ {./Figures/}, {./Section 2/}, {./Section 5/}, {./Section 6/}} 
\usepackage{pgfplots}
\pgfplotsset{compat=1.18}

\title{Topolons: Stable Particle-Like Remnants of Collapsed Vacuum Bubbles in a Three-Form Gauge Sector}

\author[1,2]{Muhammad Ghulam Khuwajah Khan\thanks{\href{mailto:b24bs1234@iitj.ac.in}{b24bs1234@iitj.ac.in}}\thanks{\href{khanmuhammadghulam@rjcollege.edu.in}{khanmuhammadghulam@rjcollege.edu.in}}\thanks{\href{dr.muhammad.ghulam.khuwajah@gmail.com}{dr.muhammad.ghulam.khuwajah@gmail.com}}}

\affil[1]{\textit{Department of Physics, Ramniranjan Jhunjhunwala College, Mumbai 400\,086, Maharashtra, India} \vspace{0.8em}}

\affil[2]{\textit{School of Artificial Intelligence and Data Science, Indian Institute of Technology Jodhpur, Jodhpur 342\,037, Rajasthan, India}}

\date{}

\begin{document}
	
\maketitle

\begin{abstract}
	We study a three-form gauge sector in four spacetime dimensions coupled to electrically charged spherical membranes whose worldvolume dynamics are governed by a Dirac--Born--Infeld action. The associated four-form field strength has no local propagating degrees of freedom and contributes a branch-dependent vacuum energy. Motivated by the Hartle--Hawking--Wu selection argument, we restrict attention to the semiclassically admissible four-form flux window for which the Hartle-Hawking wave function has support. We then endow the bubble wall with a worldvolume $U(1)$ gauge field carrying quantized monopole flux $n \in \mathbb{Z}$ and evaluate the full DBI energy of the resulting spherical configurations. We show that the energetically preferred branch collapses toward a microscopic core rather than stabilizing at finite radius. Furthermore, we show that if the bubble carries a nonzero monopole flux, the energy does not vanish in the collapsed limit. Instead, the bubble relaxes to a finite-energy remnant whose mass is set by the wall scale and the conserved flux. We interpret these objects as stable flux-supported particle-like states, which we call topolons. Within the admissible sector, the effective energy analysis distinguishes stable collapsed remnants from the contrasting runaway vacuum-decay channel, thereby isolating the sector relevant for cosmological relic formation. At macroscopic distances, topolons behave as heavy localized states and provide a concrete microphysical realization of a dark relic candidate. The detailed cosmological abundance and phenomenology are left for future work. More broadly, the present construction supplies the nonperturbative	remnant component of an economical three-form framework in which the
	homogeneous four-form flux may neutralize the dominant permanent
	vacuum-energy contribution, while a dynamical scalar supplies the
	temporary positive energy required for a no-boundary inflationary
	history. Additional four-form dynamics, another sector, or presently
	unknown physics may leave the small residual vacuum energy observed at
	late times. The present paper does not claim a complete unified
	cosmology, but establishes the microscopic topolon sector required for
	such a program.
\end{abstract}
\noindent \textbf{\textit{keywords:}} Topolons; Three-form gauge field; Four-form flux; Dirac–Born–Infeld membrane; Vacuum bubbles; Dark relics

\newpage

\section{\centering Introduction}

The idea that an antisymmetric three-form gauge potential $C_{3}$ with components $C_{\mu\nu\rho}$ may play a central role in the cosmological constant problem goes back to the early 1980s. In four spacetime dimensions, the associated four-form field strength,

\begin{equation}
	F_{\mu\nu\rho\sigma} = 4 \, \partial_{[\mu} C_{\nu\rho\sigma]} \, ,
\end{equation}
has no local propagating degrees of freedom. On shell, it is a spacetime constant and therefore contributes to the stress-energy tensor exactly as a vacuum energy term. Early work by Aurilia, Nicolai and Townsend, and by Duff and van Nieuwenhuizen, made clear that such three-form gauge sectors with four-form field strengths can arise naturally and can act effectively as a cosmological constant contribution \cite{AuriliaNicolaiTownsend1980, DuffvN1980}.
\\ \\
This observation soon became intertwined with attempts to understand why
the observed cosmological constant is so small. The no-boundary proposal of
Hartle and Hawking assigns a semiclassical Euclidean wavefunction to the
Universe and, for a positive de Sitter saddle, exponentially enhances
histories with smaller positive values of the cosmological constant
\cite{HH1983}. The cosmological constant was still a fixed
parameter in that original construction. Building on this no-boundary
weighting and on Baum's related Euclidean argument, Hawking subsequently
introduced a three-form gauge potential $C_3$, whose on-shell four-form
field strength $F_4 = dC_3$ contributes an integration constant $q$ to
the effective cosmological constant \cite{Baum1983, Hawking1984}. The
different values of $q$ then label distinct four-form flux branches, and
the Hartle--Hawking weighting formally favors the admissible branch for
which the positive effective cosmological constant approaches zero from
above.
\\ \\
Hawking's original implementation was later challenged by Duff, who showed
that varying the metric before imposing the on-shell four-form solution
changes the sign of the flux contribution to the effective cosmological
constant \cite{Duff1989}. Wu subsequently argued that the apparent
discrepancy is resolved by formulating the no-boundary wavefunction in the
fixed-flux representation and including the corresponding Legendre boundary
term \cite{Wu2008}. In the resulting Hartle--Hawking--Duff--Wu picture
adopted in the present work, the constant four-form flux labels a family of
vacuum branches, while the no-boundary weighting selects the branch with
the smallest attainable positive value of $\Lambda_{\rm eff}$. The
exactly neutral configuration, $\Lambda_{\rm eff} = 0$, is understood as
the limiting endpoint of this semiclassically admissible family.
\\ \\
Henneaux and Teitelboim further clarified the canonical structure of
three-form gauge theories by showing how the cosmological constant may be
treated as a dynamical variable conjugate to a three-form degree of freedom
\cite{HenneauxTeitelboim1984}. Brown and Teitelboim then coupled the
three-form gauge potential to electrically charged membranes, whose
nucleation produces discrete jumps in the four-form flux and therefore
mediates transitions between different vacuum-energy branches
\cite{BT1987,BT1988}.
\\ \\
Related ideas have continued to appear in several modern approaches to vacuum energy. In $q$-theory, the quantum vacuum is modeled as a self-sustained medium characterized by a conserved variable $q$, which can be realized through a three-form gauge field and whose equilibrium value relaxes the vacuum pressure toward zero \cite{KV2008, KV2009, VolovikBook}. In string compactifications, multiple four-form fluxes generate a dense discretuum of effective cosmological constants, making small positive $\Lambda$ possible as one choice in a large landscape \cite{BoussoPolchinski2000}. Four-form sectors also play a central role in vacuum energy sequestering, where global or local constraints enforced by four-forms remove matter vacuum energy from the gravitational field equations \cite{KaloperPadilla2014PRL, KaloperPadilla2014PRD, KaloperPadilla2015PRL, KaloperPadilla2016PRL, KaloperPadilla2017PRL, Avelino2014PRD, BenDayan2016JCAP, DAmicoEtAl2017}. More broadly, renewed interest in vacuum energy and late-time acceleration has also been sharpened by current large-scale-structure data, including recent DESI results for dynamical dark energy and their interpretation \cite{DESI2024VI, DESIDR1, DESIDR2, KhanTsDE, Khan2025}.
\\ \\
In addition to this, inflation has also been incorporated into the no-boundary framework in several different ways. The original Hartle--Hawking proposal already allowed matter fields to be included in the Euclidean path integral, while Halliwell and Hawking developed the treatment of scalar and metric perturbations on inflationary backgrounds, thereby connecting the no-boundary state with the primordial fluctuations responsible for large-scale structure \cite{HH1983, HalliwellHawking1985}. Hawking and Turok later constructed open inflationary histories from no-boundary instantons without requiring a false vacuum, although the corresponding Euclidean saddles contain a controlled singularity \cite{HawkingTurok1998}. Turok and Hawking extended this construction by including a four-form field strength, allowing the effective cosmological constant to vary alongside the inflationary sector \cite{TurokHawking1998}. Bousso and Chamblin subsequently showed that the singularity could be replaced by a charged membrane, yielding nonsingular instantons describing the creation of an open inflationary universe \cite{BoussoChamblin1999}.
\\ \\
A complementary development was given by Hartle, Hawking and Hertog, who
studied generally complex no-boundary saddles containing a dynamical scalar
field and identified the classicality conditions under which they continue
to Lorentzian inflationary histories. They also investigated the probability
measure over such histories, including volume weighting and the regime of
eternal inflation \cite{HartleHawkingHertog2008Classical, 	HartleHawkingHertog2008Measure, HartleHawkingHertog2010}. More recently,
Maldacena provided an analytic treatment of the no-boundary geometry in
slow-roll inflation and emphasized the associated measure and spatial-curvature issues, while a comprehensive account of these developments and their open questions is given in Ref.~\cite{Lehners2023}
\cite{Maldacena2024}. These works demonstrate that inflationary evolution can in principle be accommodated within the no-boundary framework. It should be mentioned here that the protection of the dynamical inflaton energy from further flux neutralization on the saddle has not been fully clarified. In future work, we plan to present a minimal construction in which flux quantization locks the system onto a vacuum-neutralizing branch, while a dynamical scalar generates a regular no-boundary saddle and the subsequent inflationary phase. This work will also shed more light and provide a more rigorous justification for the existence of an admissible sector for the four-flux $q$ presented in this work. 
\\ \\
Taken together, these developments suggest that a three-form gauge sector
with four-form flux can provide an unusually economical organizing
structure for a broader cosmological framework. At the homogeneous level,
the four-form flux labels vacuum branches and, within the
Hartle--Hawking--Duff--Wu construction, may select an endpoint at which the
dominant permanent vacuum-energy contribution is neutralized. When the
neutral branch is supplemented by a dynamical scalar field, the scalar may
supply the temporary positive energy required for a no-boundary
inflationary history without restoring the cancelled permanent vacuum
contribution. Related four-form constructions, including \(q\)-theory,
vacuum-energy sequestering, and flux-discretuum models, further indicate
that a small residual late-time vacuum energy may survive through
additional dynamics or additional sectors.
\\ \\
In the present work, however, we move beyond the homogeneous background
of the three-form gauge sector and study its nonperturbative
charged-membrane configurations and their worldvolume dynamics. Our
central question is whether a vacuum bubble interpolating between
distinct four-form flux branches must collapse into a trivial
zero-energy configuration, or whether nonlinear worldvolume physics can
produce a localized finite-energy remnant. We show that, when the
membrane is governed by a Dirac--Born--Infeld (DBI) action and carries a
conserved worldvolume \(U(1)\) monopole flux, the collapsed endpoint
need not be trivial. For the physically admissible branch transitions selected by the Hartle--Hawking--Wu condition, the bubble can instead
approach a microscopic finite-energy configuration supported by
topology, flux conservation, and nonlinear brane dynamics. We refer to
these particle-like remnants as \emph{topolons}.
\\ \\
The resulting picture is economical. At the homogeneous level, the constant four-form flux labels vacuum branches and contributes a branch-dependent vacuum energy, while membranes electrically charged under the three-form potential mediate discrete transitions between these branches. At the nonperturbative level, the same membranes may carry an independent quantized worldvolume monopole flux, allowing their collapse to terminate in finite-energy, particle-like remnants. The aim of the present paper is to establish the existence, energetic stability, and classification of these objects within the effective DBI description. Their possible production mechanisms, relic abundance, effective core size, gravitational backreaction, and viability as dark-matter candidates are deferred to a companion cosmological study.
\\ \\
More broadly, these developments in literature suggests that a single three-form gauge sector may provide a common theoretical backbone for several otherwise distinct aspects of cosmology. Its homogeneous four-form flux may participate in neutralizing the dominant permanent vacuum-energy contribution; additional dynamics may leave a small residual late-time
vacuum energy; and now the present work demonstrates that its nonperturbative charged-membrane configurations may furnish a dark-matter candidate (topolons). A dynamical scalar field may, in turn, supply the temporary positive energy required for an inflationary no-boundary history without restoring the neutralized permanent vacuum contribution. These results do not yet constitute a complete model of the dark sector, but they motivate a unified program in which vacuum energy, inflationary initial conditions, dark energy, and dark matter all can be explained minimally with a only a single ingredient, namely the universe appended with a three-form gauge sector.
\\ \\
The remainder of the paper is organized as follows. We first review the three-form framework, its stress-energy structure, and its coupling to charged membranes. We then analyze the branch structure and the semiclassically admissible transitions selected by the Hartle--Hawking--Wu criterion. Building on this setup, we study the DBI energy functional for charged spherical bubbles with conserved worldvolume monopole flux and show that the collapse endpoint can be a finite-mass remnant rather than a trivial vacuum state. We then interpret and classify these remnants, discuss their regime of validity and limitations as an effective
description, and comment on their possible relevance as cosmological relics. The conventions used throughout are summarized in Table 1.
\\ \\
In this work, we adopt the following conventions,

\begin{table}[h!]
	\centering
	\begin{tabular}{ll}
		\hline
		\textbf{Convention} & \textbf{Definition} \\
		\hline
		Units & $\hbar = c = k_B = 1$. \\
		Gravitational constant & $G = 1/M_{\mathrm{Pl}}^2$ \\
		Planck mass & $M_{\mathrm{Pl}} = 1.22 \times 10^{19}\,\mathrm{GeV}$ \\
		Lorentzian signature & $(-,+,+,+)$ \\
		Euclidean signature & $(+,+,+,+)$ \\
		Levi-Civita symbol & $\tilde{\epsilon}_{\mu\nu\rho\sigma}$ and $\tilde{\epsilon}^{0123} = -1$ \\
		Covariant Levi-Civita tensor & $\epsilon_{\mu\nu\rho\sigma} = \sqrt{-g}\,\tilde{\epsilon}_{\mu\nu\rho\sigma}$ \\
		Contravariant Levi-Civita tensor & $\epsilon^{\mu\nu\rho\sigma} = \dfrac{1}{\sqrt{-g}}\,\tilde{\epsilon}^{\mu\nu\rho\sigma}$ \\
		\hline
	\end{tabular}
	\caption{Summary of conventions.}
	\label{tab:summary_conventions}
\end{table}
\FloatBarrier

\section{\centering The Three-Form Gauge Sector $C_3$}

\subsection{\centering Equation of Motion and the Stress Energy Tensor}

The three-form gauge potential is given by,

\begin{equation}
	C_{3} \; = \; \frac{1}{3!} \, C_{\mu\nu\rho} \, dx^{\mu} \wedge dx^{\nu} \wedge dx^{\rho} \, ,
\end{equation}
with four-form field strength,

\begin{equation}
	F_{4} \; = \; dC_{3} \, , \qquad F_{\mu\nu\rho\sigma} \; = \; 4 \, \partial_{[\mu} C_{\nu\rho\sigma]} \, .
\end{equation}
Given that the Levi-Civita connection is torsion free (symmetric in its lower indices), it can be shown that the exterior derivative is equal to the antisymmetrized covariant derivative (see Sec.~3, discussion leading up to equation (3.29) in \cite{carrollnotes}), that is,

\begin{equation}
	\partial_{[\mu} C_{\nu\rho\sigma]} \, = \, \nabla_{[\mu} C_{\nu\rho\sigma]} \, .
	\label{Nabla Partial Commute}
\end{equation}
As a result, one may freely swap $\partial$ and $\nabla$ inside any totally antisymmetric combination without changing the result. The dynamics of $C_3$ follows from the standard Maxwell-like action and can be expressed as,

\begin{equation}
	S_3 \, = \, - \, \frac{1}{2 \cdot 4!} \, \int d^4 x \, \sqrt{-g} \, F_{\mu\nu\rho\sigma} F^{\mu\nu\rho\sigma} \, .
	\label{S3 equation 1}
\end{equation}
Varying $S_3$ with respect to $C_{\nu\rho\sigma}$ and using \eqref{Nabla Partial Commute} gives,

\begin{align}
	\delta S_3 & = - \, \frac{1}{3!} \, \int d^4 x \, \sqrt{-g} \, F^{\mu\nu\rho\sigma} \, \partial_{\mu} \delta C_{\nu\rho\sigma}
	\nonumber \\
	& = - \, \frac{1}{3!} \, \int d^4 x \, \sqrt{-g} \, F^{\mu\nu\rho\sigma} \, \nabla_{\mu} \delta C_{\nu\rho\sigma} \, .
\end{align}
Integrating by parts and discarding the surface term leads to,

\begin{equation}
	\delta S_3 \, = \, \frac{1}{3!} \, \int d^4 x \, \sqrt{-g} \, \big( \nabla_{\mu} F^{\mu\nu\rho\sigma} \big) \, \delta C_{\nu\rho\sigma} \, .
\end{equation}
Requiring $\delta S_3 \, = \, 0$ for arbitrary $\delta C_{\nu\rho\sigma}$ gives the equation of motion for the three-form gauge potential	as,

\begin{equation}
	\nabla_{\mu} F^{\mu\nu\rho\sigma} \, = \, 0 \, .
\end{equation}
The quantity $\nabla_{\mu} F^{\mu\nu\rho\sigma}$ can be expressed as (see Sec.~3 in \cite{carrollnotes}),

\begin{equation}
	\nabla_{\mu} F^{\mu\nu\rho\sigma} \, = \, \frac{1}{\sqrt{-g}} \, \partial_{\mu} \big( \sqrt{-g} \, F^{\mu\nu\rho\sigma} \big) \, ,
\end{equation}
and the equation of motion therefore becomes,

\begin{equation}
	\partial_{\mu} \big( \sqrt{-g} \, F^{\mu\nu\rho\sigma} \big) \, = \, 0 \, ,
\end{equation}
which has the general solution,

\begin{equation}
	F^{\mu\nu\rho\sigma} \, = \, \frac{q}{\sqrt{-g}} \, \tilde{\epsilon}^{\mu\nu\rho\sigma} \, = \, q \, \epsilon^{\mu\nu\rho\sigma} \, ,
\end{equation}
where $q$ is constant on each connected source-free spacetime region
and has dimension $[q]={\rm mass}^2$ in natural units. In the presence
of charged membranes, $q$ may take different constant values on the
two sides of the membrane. With this, note that equation \eqref{S3 equation 1} can now be expressed in terms of $q$ as,

\begin{equation}
	S_3 \, = \, - \, \frac{1}{2 \cdot 4!} \, \int d^4 x \, \sqrt{-g} \, q^2 \, \epsilon_{\mu\nu\rho\sigma} \, \epsilon^{\mu\nu\rho\sigma} \, .
\end{equation}
The contraction of the volume form with itself gives, $\epsilon_{\mu\nu\rho\sigma} \, \epsilon^{\mu\nu\rho\sigma} \, = \, -4!$ and therefore, the action $S_3$ reduces to,

\begin{equation}
	S_3 \, = \, \frac{1}{2} \, \int d^4 x \, \sqrt{-g} \, q^2 \, .
\end{equation}
Using the equation above, one can show that the stress-energy tensor of the three-form gauge sector, evaluated on the four-form flux solution, is (see Appendix A),

\begin{equation}
	T^{(q)}_{\mu\nu} \; = \; -\frac{1}{2} \, q^{2} \, g_{\mu\nu} \, .
	\label{eq:stress_energy_tensor_of_q}
\end{equation}
The stress-energy tensor of the four-form sector as written above, is exactly of the form of a fluid with equation of state $p_q \, = \, - \rho_q$ and therefore, the $q$ sector behaves exactly like a Cosmological Constant. Furthermore, the stress-energy tensor of $C_3$ signifies that different values of $q$ label topologically distinct vacuum states of the universe. A transition from one flux branch $q$ to another branch $q'$ requires a charged source, such as the nucleation of a membrane electrically coupled to \(C_3\), which produces a discrete jump in the four-form flux. Depending on the relative branch energies, such a process may describe vacuum decay or a transition between degenerate flux branches \cite{BT1987, BT1988, CDL1980}.

\subsection{\centering Electrically Charged Membranes and the Flux Jump Condition}

\begin{figure}[htbp]
	\centering
	\includegraphics[width=\textwidth]{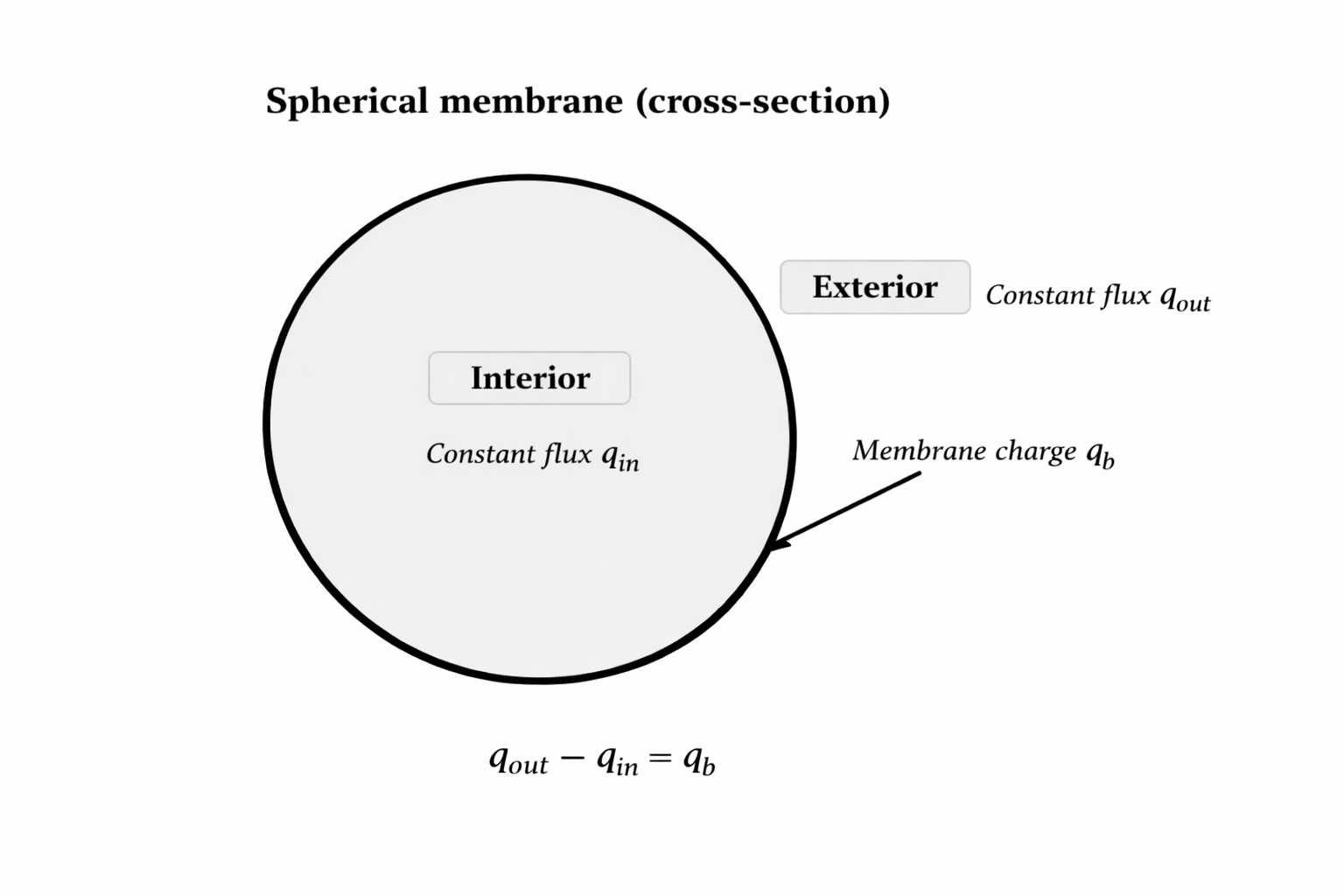}
	\caption{Spherical membrane separating two constant four-form flux sectors. The interior and exterior are characterized by constant values $q_{\rm in}$ and $q_{\rm out}$. The membrane carries charge $q_{b}$, enforcing the jump condition $q_{\rm out} - q_{\rm in} = q_{b}$.}
	\label{fig:vacuum_bubble_jump}
\end{figure}

It is known that an object with $(p - 1)$ spatial dimensions couples electrically to a $p$-form potential \cite{PolchinskiBook2}. In our case $p \, = \, 3$, so the degrees of freedom that couple to $C_3$ are extended two-dimensional objects, commonly referred to as 2-branes. We focus on the simplest symmetry class that is both physically motivated and technically convenient, namely a closed spherical membrane with worldvolume $\Sigma^{3} = \mathbb{R} \times S^{2}$. The configuration is therefore fully characterized by a single collective coordinate, the physical radius $R$. We do not promote $R$ to a dynamical worldvolume degree of freedom to describe radial fluctuations or motion. This is because in later Sections, we show that in the effective DBI description, the energetically preferred configuration corresponds to a collapsed remnant, for which the radius approaches $R \rightarrow 0$ up to microscopic cutoff effects. At energies well below this scale the dynamics of radial excitations decouples from the low energy spectrum, and introducing an independent degree of freedom associated with $R$ is therefore neither required nor physically motivated. 
\\ \\
The action for the three-form sector coupled to a membrane of charge $q_{b}$ is given by,

\begin{equation}
	S_{3} = -\frac{1}{2 \cdot 4!} \int d^{4}x \, \sqrt{-g} \; F_{\mu\nu\rho\sigma} F^{\mu\nu\rho\sigma} \; + \; q_{b} \int_{\Sigma^{3}} C_{3} \, .
	\label{eq:S3_with_source}
\end{equation}
Here, the term $\displaystyle q_{b} \int_{\Sigma^{3}} C_{3}$ represents the electric coupling between the three-form gauge potential $C_{3}$ and the source, namely the 2-brane whose worldvolume is $\Sigma^{3}$. If the source is spherical, then in four spacetime dimensions, the four-form flux $q$ obeys a jump condition across the membrane (consult Appendix B for a detailed derivation) which reads,

\begin{equation}
	q_{\mathrm{out}} - q_{\mathrm{in}} \; = \; q_{b} \, ,
	\label{eq:flux_jump}
\end{equation}
where $q_{\rm out}$ is the value of the four-form flux outside the
bubble, while $q_{\rm in}$ is its value inside the bubble. Thus, the membrane interpolates between two constant-flux sectors, or equivalently, the four-form flux $q$ undergoes a discrete jump across the membrane because the membrane is electrically charged under the three-form gauge potential $C_3$. A schematic diagram of the membrane-induced jump is shown in Fig.~\ref{fig:vacuum_bubble_jump}.

\FloatBarrier

\noindent In the next section, we discuss the role of the four-form flux sector in the context of the Hartle--Hawking (HH) wave function of the universe. In particular, we argue that the HH weighting restricts the allowed range of $q$ to the subset of parameter space for which a regular compact Euclidean cap exists.

\section{\centering Wave-Function of the Universe}
\label{sec:HHW}
In order to obtain the likelihood of finding a Hubble patch with initial 3-geometry $h_{ij}$ and matter data $\Phi_0$ which then evolves classically, the `no boundary proposal' was put forth by Hartle and Hawking \cite{HH1983}. The demand is that instead of having a sharp edge (`initial singularity'), the history of the universe rounds off smoothly in Euclidean time. Topologically, the beginning looks like a smooth 4 sphere $S^4$, which caps off the universe's past and is aptly referred to as the `Euclidean Cap'. Under this scheme, it is possible to compute the ground state wavefunction $\Psi_0$ of the universe via a Euclidean path integral over compact, boundary-less four geometries $g$ and matter fields $\Phi$ that are (i) regular at the south pole of the $S^4$ cap $(\tau \, = \, 0)$ and (ii) match a prescribed induced three metric $h_{ij}$ and matter data $\Phi_0$ on the nucleation surface $\Sigma_{\ast}$, which for the standard saddle is the equator of the $S^4$ cap. Therefore, the Hartle-Hawking wavefunction for the ground state of the universe $\Psi_0$ can be expressed as,

\begin{equation}
	\Psi_0[h_{ij}, \Phi_0] = \int_{g|_{\partial M \, = \, h} \, , \, \Phi|_{\partial M \, = \, \Phi_0}} \mathcal{D}g \, \mathcal{D}\Phi \, \exp(- S_E[g,\Phi]) \, ,
\end{equation}
where $S_E$ is the Euclidean action. If we consider only gravity, then $S_E$ reduces to,

\begin{equation}
	S_E[g] = - \, \frac{1}{16 \pi G} \, \int d^4 x \, \sqrt{g} \, (R - 2\Lambda) \, .
	\label{Euclidean Action HH}
\end{equation}
For the Euclidean de Sitter saddle, one has $R = 4 \Lambda$ and $a = H^{-1} = \sqrt{3/\Lambda}$. The no-boundary saddle contributing to the wavefunction is not the complete four-sphere, but the regular Euclidean hemisphere bounded by the nucleation surface at the equator, that is the Euclidean cap. Consequently, equation \eqref{Euclidean Action HH} becomes,

\begin{equation}
	S_E[g] \; = \; - \, \frac{\Lambda}{8\pi G} \int_{\frac{1}{2} \, S^4} d^4x \, \sqrt{g} \; = \; - \, \frac{\Lambda}{8\pi G} \, {\rm Vol} \left(\frac{1}{2} \, S^4\right).
	\label{HH2}
\end{equation}
The volume of a complete four-sphere of radius \(a\) is,
\footnote{
	Given a Cosmological Constant $\Lambda$, it is important to note a compact Euclidean cap exists only if $\Lambda > 0$, in which case the cap is a four sphere $S^4$ with radius $a \, = \, H^{-1} \, = \, \sqrt{3/\Lambda}$. For $\Lambda = 0$, the cap geometry is $\mathbb{R}^4$ and for $\Lambda < 0$, the cap is $H^4$ (the hyperbolic space) both of which are non-compact. In essence, the Hartle-Hawking (HH) wavefunction is well defined only for the class of de Sitter universes. For details on the compactness criteria see \cite{HH1983, Halliwell1989, Spradlin2001}.
}

\begin{equation}
	{\rm Vol}(S^4) = \frac{8\pi^2}{3} \, a^4 = \frac{24 \, \pi^2}{\Lambda^2}.
\end{equation}
Therefore, the volume of the hemispherical Euclidean cap is,

\begin{equation}
	{\rm Vol} \left(\frac{1}{2}S^4\right) \; = \; \frac{1}{2} \, {\rm Vol}(S^4) \; = \; \frac{12 \, \pi^2}{\Lambda^2}.
\end{equation}
Substituting this result into equation \eqref{HH2} gives,

\begin{equation}
	S_E[g] \; = \; - \left(\frac{\Lambda}{8\pi G}\right) \left(\frac{12 \, \pi^2}{\Lambda^2}\right) \; = \; - \, \frac{3 \, \pi}{2 \, G \, \Lambda} \, .
	\label{HH3}
\end{equation}
The Hartle-Hawking tunneling amplitude $\Psi_{HH}$ for this saddle is then,

\begin{equation}
	\Psi_{HH} \, \propto \, e^{-S_E[g]} \, \propto \, \exp \left( \frac{3 \, \pi}{2\, G \, \Lambda} \right) \, .
	\label{HH saddle weight}
\end{equation}
The corresponding probability weight $P_{HH}$ is obtained by squaring the
wavefunction amplitude,

\begin{equation}
	P_{\rm HH} \; \propto \; \left|\Psi_{\rm HH}\right|^2 \; \propto \; \exp \left(\frac{3 \, \pi}{G \, \Lambda}\right) \, .
	\label{HH saddle weight_2}
\end{equation}
It is easy to see that the Hartle--Hawking weight exponentially favors the smallest attainable positive $\Lambda$ and it grows without bounds as $\Lambda \to 0^+$. Note that $\Lambda$ is not a dynamical variable but rather is a parameter of the theory and the Hartle-Hawking saddle \eqref{HH saddle weight} clearly states that if $\Lambda$ in principle could vary, then the saddle with smallest positive $\Lambda$ would dominate. Motivated to make $\Lambda$ dynamical, Hawking \cite{Hawking1984} augmented the gravitational sector by a three form potential $C_3$ along with its four-form field strength $F_4$ exactly as defined in Sec.~2. This allowed Hawking to perform a path integral over the distinct four-form flux sectors (different $q$) and dynamically select the smallest attainable positive $\Lambda$. Hawking’s initial 1984 calculation was shown by Duff \cite{Duff1989} to be incorrect and was later corrected by Wu \cite{Wu2008}.
\\ \\
To allow an effective vacuum energy that can scan between flux sectors, Hawking \cite{Hawking1984} supplemented gravity by a three-form potential $C_3$ with its four-form field strength $F_4 = dC_3$ and Euclidean action,

\begin{equation}
	S_E[g, C] \; = \; -\int d^4x\,\sqrt{g} \, \left[ \frac{1}{16\pi G}(R - 2\Lambda_0) - \frac{1}{48} \, F_{\mu\nu\rho\sigma} \, F^{\mu\nu\rho\sigma} \right] \, , \qquad F_4 = dC_3 \, .
	\label{eq:SE_Hawking}
\end{equation}
Hawking's original 1984 procedure (integrating out $C_3$ at fixed boundary potential $C_3|_{\Sigma_\ast}$ and substituting back into the action) yields an Einstein equation of the form (see Appendix C.1 for a detailed derivation),

\begin{equation}
	G_{\mu\nu} + \Lambda_{\rm eff} \, g_{\mu\nu} \; = \; 0 \, , \qquad \Lambda_{\rm eff} \; = \; \Lambda_0 + 4\pi G \, q^2 \, ,
	\label{eq:LambdaEff_Hawking}
\end{equation}
The Hartle-Hawking tunneling weight reads,

\begin{equation}
	|\Psi_q| \; \propto \; \exp \left( \frac{3 \, \pi}{2 \, G\Lambda_{\rm eff}} \right) \; \propto \; \exp \left( \frac{3 \, \pi}{2 \, G \, (\Lambda_0 + 4\pi G q^2)} \right)
\end{equation}
Note that if $\Lambda_0 > 0$, the exponent decreases with $|q|$ and therefore the dominant sector is $q \, = \, 0$. If on the other hand $\Lambda_0 < 0$, only $|q| > \sqrt{-\Lambda_0/(4\pi G)}$ gives $\Lambda_{\rm eff} > 0$ (so an $S^4$ cap exists) and among such $|q|$'s, the smallest allowed is chosen so that $\Lambda_{\rm eff}$ is as small as possible from above and the weight is maximized. This is the manner in which Hawking advertised how HH favors small positive $\Lambda$ at birth.
\\ \\
Duff \cite{Duff1989} pointed out that for constrained systems the operations ``substitute the on-shell value'' and ``vary the metric'' need not commute. Varying the $F_4$ field-strength term with respect
to $g_{\mu\nu}$ before substituting the on-shell flux solution
$F_4 = q \, {\rm vol}_4$ gives instead (see Appendix C.2),

\begin{equation}
	G_{\mu\nu} + (\Lambda_0 - 4\pi G \, q^2) \, g_{\mu\nu} = 0 \, , \qquad
	\Lambda_{\rm eff} \; = \; \Lambda_0 - 4\pi G \, q^2 \, ,
	\label{eq:LambdaEff_Duff}
\end{equation}
which differs from equation \eqref{eq:LambdaEff_Hawking} in the sign of
the four-form flux contribution to the effective cosmological constant. Importantly, Hawking’s qualitative conclusion (that HH favors the smallest positive $\Lambda_{\rm eff}$ among the allowed flux $(q)$ sectors) survives; however is now realized by choosing that $q$ which minimizes $\Lambda_{\rm eff} \, = \, \Lambda_0 - 4\pi G q^2$. In essence, the Hartle-Hawking tunneling weight now reads,

\begin{equation}
	|\Psi_q| \; \propto \; \exp \left( \frac{3 \, \pi}{2\, G \, \Lambda_{\rm eff}} \right) \; \propto \; \exp \left( \frac{3 \, \pi}{2 \, G \, (\Lambda_0 - 4\pi G q^2)} \right)
\end{equation} 
Consequently, a compact Euclidean cap exists (absent other positive contributions) only if $\Lambda_{\rm eff} > 0$, which requires $\Lambda_0 > 0$.
\\ \\
Finally, Wu \cite{Wu2008} showed that the sign issue is resolved by adopting the correct boundary data at $\Sigma_\ast$. The appropriate variational problem fixes the flux (equivalently $\star F$) on the nucleation surface, not the potential $C_3$. This is implemented by adding the Legendre boundary term,

\begin{equation}
	S_E[g, C] \; \longrightarrow \; S_E[g, C] \; - \; \int_{\Sigma_\ast} C\wedge \star F \, ,
	\label{eq:LegendreTerm}
\end{equation}
after which, the operations of first integrating out $C_3$ and then varying $g_{\mu\nu}$ or vice-versa then commute and one consistently obtains the Duff result \eqref{eq:LambdaEff_Duff} (a comprehensive derivation is included in Appendix C.3).
\\ \\
In the remainder of the paper, we use \eqref{HH saddle weight_2} together with the corrected flux-dependent effective cosmological constant \eqref{eq:LambdaEff_Duff} to motivate that, among the flux sectors admitting a compact Euclidean cap ($\Lambda_{\rm eff} > 0$), the Hartle--Hawking weight favors those with the smallest positive $\Lambda_{\rm eff}$. In the next section we study the vacuum energy as a function of $q$ and show that it vanishes at two degenerate endpoint values $q = \pm q_0$, while being larger for intermediate $q \in [- q_0, + q_0]$.
\\ \\
We have therefore arrived at the corrected Hawking--Duff--Wu picture which is that in four spacetime dimensions, the three-form gauge sector carries no
local propagating bulk degrees of freedom, while the constant flux of its four-form field strength labels distinct vacuum branches through $\Lambda_{\rm eff}(q)$. In the next section, we identify the semiclassically admissible range of these flux branches and isolate the
endpoint values at which the permanent vacuum-energy contribution is
neutralized.

\section{\centering The Admissible Four-Form Flux Sector}

From equation \eqref{eq:LambdaEff_Duff}, the effective cosmological constant on a branch labeled by the constant four-form flux \(q\) is,

\begin{equation}
	\Lambda_{\rm eff}(q) = \Lambda_0 - 4\pi G \, q^2 \, .
	\label{Effective Lambda}
\end{equation}
In the Hartle--Hawking--Wu picture the nucleation probability for a compact $S^4$ universe is nonzero only if $\Lambda_{\rm eff} > 0$, and within this range the probability is largest when $\Lambda_{\rm eff}$ approaches zero from above. In order to obtain the value of the four–flux $q$ at which the effective cosmological constant vanishes, we solve for $\Lambda_{\rm eff}(q) = 0$, which gives,

\begin{equation}
	q^2 = \frac{\Lambda_0}{4\pi G} \qquad \Longrightarrow \qquad q = \pm \sqrt{\frac{\Lambda_0}{4\pi G}} \, .
	\label{q0 value}
\end{equation}
Let,

\begin{equation}
	q_0 \, \equiv \; \sqrt{\frac{\Lambda_0}{4\pi G}}
\end{equation}
Thus from equation \eqref{q0 value}, we see that there are two flux values, $+q_0$ and $-q_0$, for which the net cosmological constant vanishes, \footnote{This degeneracy reflects the fact that the stress-energy tensor generated by the on-shell four-form field strength depends only	on $q^2$, $T_{\mu\nu}^{(q)} \propto q^2 \, g_{\mu\nu}$, so the sign of the four-form flux $q$ does not affect the vacuum-energy density.} that is,

\begin{equation}
	\Lambda_{\rm eff}(\pm q_0) = 0 \, .
	\label{Vanishing Lambda}
\end{equation}
For $|q| < |q_0|$ or equivalently, for $-q_0 < q < q_0$, the effective cosmological constant is positive,

\begin{equation}
	|q| < |q_0| \quad \Longrightarrow \quad \Lambda_{\rm eff}(q) > 0
	\label{feasible condition}
\end{equation}
and the Euclidean $S^4$ saddle exists. In this regime the Hartle--Hawking no–boundary wave function is defined and the usual argument applies, namely that configurations with smaller $\Lambda_{\rm eff}$ are exponentially preferred. 
\\ \\
On the other hand, for $|q| > |q_0|$, or equivalently $q > q_0$ or $q < - q_0$, the effective cosmological constant becomes negative,

\begin{equation}
	|q| > |q_0| \quad \Longrightarrow \quad \Lambda_{\rm eff}(q) < 0  
\end{equation}
and a compact $S^4$ solution is absent. Therefore, in the simplest implementation of the no–boundary proposal we restrict attention to the condition \eqref{feasible condition} or the interval,

\begin{equation}
	q \in [-q_0, \, +q_0] \, ,
\end{equation}
within which the total effective cosmological constant remains positive
and the compact \(S^4\) saddle exists.
\\ \\
It should be emphasized that the local equations of motion of the
three-form gauge sector do not, by themselves, terminate the algebraic
flux-branch structure at a finite value of $|q|$. Rather, we impose a restriction on the \emph{admissible} flux sector, understood as a consistency condition on the effective theory together with the cosmological state that prepares its semiclassical branches. Concretely, the condition \eqref{feasible condition} is defined so that all admissible branches satisfy $\Lambda_{\rm eff}(q)\ge 0$.
\\ \\
In particular, the motivation for \eqref{feasible condition} comes from the fact that a homogeneous vacuum branch contributes semiclassically only if it admits a regular compact Euclidean saddle. For $\Lambda_{\rm eff}(q) > 0$, the relevant saddle is the round Euclidean de Sitter cap, namely an $S^4$ of radius,

\begin{equation}
	a = H^{-1} = \sqrt{\frac{3}{\Lambda_{\rm eff}(q)}} \, .
\end{equation}
By contrast, for $\Lambda_{\rm eff}(q)<0$ there is no regular compact $O(4)$-symmetric Euclidean continuation of this type, so such branches are not prepared by the no-boundary saddle at semiclassical order. In this sense, the no-boundary prescription does not merely weight different flux branches but rather it selects the subset of branches on which the semiclassical state has support.
\\ \\
Strictly speaking, the compact $S^4$ saddle exists only for $\Lambda_{\rm eff}(q) > 0$. The endpoint values $q = \pm q_0$, for which $\Lambda_{\rm eff} = 0$, should therefore be understood as the limiting closure of the admissible set, obtained as $\Lambda_{\rm eff} \to 0^+$. Equivalently, one may regulate the discussion by working with $q = \pm(q_0 - \varepsilon)$ at intermediate stages and only at the end take $\varepsilon \to 0^+$.
\\ \\
We therefore interpret \eqref{feasible condition} not as a theorem of
the local bulk dynamics alone, but as a statement about the set of
semiclassically realizable flux branches selected by the no-boundary
state and inherited by the effective theory.
\\ \\
The motivation for restricting attention to the admissible sector is, at the present stage, primarily phenomenological and structural rather than derived from a complete theory of the primordial quantum state. A more fundamental justification of the admissible flux sector will emerge as a secondary consequence in a forthcoming work devoted to building a framework for simultaneous treatment of the neutralization of the vacuum energy density and cosmic inflation, where additional conditions will be imposed on the ground-state wave function of the Universe. In particular, the no-boundary construction will be supplemented by restrictions on the allowed four-form sector and on the class of admissible Euclidean saddle configurations. These conditions will provide a more rigorous basis for selecting the flux interval adopted here. The present analysis should therefore be regarded as the effective description of this admissible sector, while its underlying cosmological selection principle will be addressed separately.
\\ \\
With this in mind, we define a distinguished bubble transition, the \emph{exact sign-flip channel}, where,

\begin{equation}
	q_{\rm out} = + q_0 \qquad \longrightarrow \qquad q_{\rm in} = - q_0 \, .
	\label{Sign flip condition}
\end{equation}
Since the effective cosmological constant depends only on $q^2$, the two endpoint branches are degenerate (condition \eqref{Vanishing Lambda}),

\begin{equation}
	\Lambda_{\rm eff}(+q_0) = \Lambda_{\rm eff}(-q_0) = 0 \, .
\end{equation}
The difference in energy density in this transition reads,

\begin{equation}
	\Delta \rho = \frac{1}{8\pi G} \left(\Lambda_{\rm eff}(q_{\rm in}) - \Lambda_{\rm eff}(q_{\rm out})\right) \, .
\end{equation}
Using equation \eqref{Effective Lambda} we obtain,

\begin{equation}
	\Delta \rho =  \frac{1}{8\pi G}((\Lambda_0 - 4 \pi G \, q_{\rm in}^2) - (\Lambda_0 - 4 \pi G \, q_{\rm out}^2)) = \frac{1}{2} (q_{\rm out}^2 - q_{\rm in}^2) \, ,
\end{equation}
and finally substituting \eqref{Sign flip condition} in the equation above, we arrive at,

\begin{equation}
	\Delta \rho = \frac{1}{2} (q_0^2 - (-q_0)^2) = 0
\end{equation}
Thus, the exact sign flip is the unique endpoint-to-endpoint transition within the admissible sector for which the interior and exterior vacuum energies are identical. This removes the volume-pressure term from the bubble energetics and makes the subsequent collapse analysis especially clean.
\\ \\
In general, starting from the endpoint parent branch $q_{\rm out} = + q_0$, the interior flux reads,

\begin{equation}
	q_{\rm in} = q_0 - q_b \, ,
\end{equation}
and remains inside the admissible interval (condition \eqref{feasible condition}) provided,

\begin{equation}
	q_{\rm min} \; \le \; q_b \; \le \; 2q_0
\end{equation}
So in general, the expression for vacuum-energy difference for the bubbles in our setup takes the form,

\begin{equation}
	\Delta \rho = \frac{1}{8 \pi G} \, \left(\Lambda_{\rm eff}(q_{\rm in}) - \Lambda_{\rm eff}(q_{\rm out})\right) = \frac{1}{2} \,\left(q_{\rm out}^{2} - q_{\rm in}^{2}\right) \, ,
\end{equation}
where $q_{\rm out} = +q _0$ and $q_{\rm in} = q_0 - q_b$, which gives,

\begin{equation}
	\Delta \rho = \frac{1}{2} \, \left(2q_0 q_b - q_b^2\right) \, .
\end{equation}
Note that,

\begin{equation}
	q_{\rm min} \le q_b\le 2q_0 \quad \Longrightarrow \quad 2 q_0 q_b \geq q_b^2
\end{equation}
and therefore,

\begin{equation}
	\Delta \rho = \frac{1}{2} \, \left(2q_0 q_b - q_b^2\right) \geq 0 \, ,
\end{equation}
that is, the energy density difference is \textit{non-negative} and the exact sign flip is recovered only for,

\begin{equation}
	q_b = 2 q_0 \qquad \Longrightarrow \qquad q_{\rm in} = - q_0 \qquad \Longrightarrow \qquad \Delta \rho = 0 \,.
\end{equation}
This distinction will be important in the stability analysis below. The exact sign-flip channel is not singled out because it is the only stable one, but because it is the cleanest one in that it removes the residual vacuum-pressure term entirely. More generic admissible jumps with $\Delta \rho > 0$ still collapse and remain stable once the DBI wall energy and conserved monopole flux are taken into account.
\\ \\
Before proceeding, it is useful to briefly discuss another possible class of bubbles, namely those with $\Delta \rho < 0$. Starting from the endpoint parent branch $q_{\rm out} = q_{0}$, the interior branch reached after nucleation is,

\begin{equation}
	q_{\rm in} \; = \; q_{0} - q_{b} \, .
\end{equation}
If the membrane charge is too large, specifically if $q_{b} > 2 q_{0}$, then one obtains

\begin{equation}
	q_{\rm in} \; = \; q_{0} - q_{b} \; < \; - q_{0} \, .
\end{equation}
Such a daughter branch lies outside the admissible interval selected by condition \eqref{feasible condition}. Equivalently, it lies in the region $q < - q_{0}$, where the semiclassical Hartle--Hawking probability weight has no support. Transitions of this type are therefore not allowed in the present setup. Hence, once condition \eqref{feasible condition} is enforced, \textit{bubbles with $\Delta \rho < 0$ cannot form}. For completeness, however, we will discuss their energetics in Sec.~6.
\\ \\
Also worth mentioning here is the fact that in a UV completion with quantized membrane charge, the allowed values of the four-form flux lie on a discrete lattice. If $q_{\min}$ denotes the elementary membrane-charge unit, that is the minimum amount by which the four-flux can jump, then a membrane may carry an integer charge,

\begin{equation}
	q_{b} \; = \; N \, q_{\min} \, , \qquad N \in \mathbb{Z}
\end{equation}
and \eqref{eq:flux_jump} corresponds to a quantized flux jump, that is,

\begin{equation}
	q_{\rm out} - q_{\rm in} \ \equiv \ \Delta q \ = \ q_b \ = \ N \, q_{\min}
\end{equation}
The \textit{most conservative assumption} is that the bubble charge is of order of the reduced Planck mass squared, $q_{\min} \sim \bar{M}_{\rm Pl}^{2}$. This is firstly because the only natural mass scale in the theory is $\bar{M}_{\rm Pl} = (8 \pi G)^{-1/2}$ and since $[q] = [\rm mass]^2$, we may write,

\begin{equation}
	q_{\rm min} \; \sim \; \bar{M}_{\rm Pl}^2 \, .
\end{equation} 
Secondly, from equation \eqref{q0 value}, one has

\begin{equation}
	q_0^2 = \frac{\Lambda_0}{4\pi G} = 2 \, \bar{M}_{\rm Pl}^2 \, \Lambda_0 \, .
\end{equation}
If the underlying vacuum-energy density is set by the Planck scale, $\rho_{\rm vac} \sim \bar M_{\rm Pl}^{\,4}$, then the corresponding
geometric cosmological constant is naturally of order $\Lambda_0 \sim \bar M_{\rm Pl}^{\,2}$, up to convention-dependent factors of order unity. It then follows that,

\begin{equation}
	q_0^2 \sim \frac{\Lambda_0}{4 \pi \, G} \sim 2 \, \bar{M}_{\rm Pl}^{4} \quad \Longrightarrow \quad q_0 \sim \sqrt{2} \, \bar{M}_{\rm Pl}^2
	\label{q0 value in terms of M}
\end{equation}
again up to factors of order unity. Thus, the four-form flux $q_0$ required
to neutralize a Planck-scale vacuum contribution is itself naturally
of order the reduced Planck scale squared. So, if the neutral endpoint is to be reached, or approached, through an order-one number of flux jumps on the Euclidean $S^4$ saddle, it is natural to consider the benchmark

\begin{equation}
	q_{\min} \; \sim \; \bar M_{\rm Pl}^{\,2} \, .
\end{equation}
This is because in such a coarse lattice, the branches nearest $q = \pm q_0$ can be accessed after only a small number of transitions and since the Hartle--Hawking--Wu weighting favors the admissible branch with the smallest positive value of \(\Lambda_{\rm eff}\), these near-endpoint lattice sectors are correspondingly preferred. The exactly neutral configuration should here be understood as the limiting endpoint \(\Lambda_{\rm eff} \to 0^{+}\); whether it is reached exactly
depends on the alignment of \(q_0\) with the quantized flux lattice.
\\ \\
Having discussed the effect of these bubbles on the bulk four-form flux
and its contribution to the vacuum energy, we now turn to the intrinsic
degrees of freedom of the membrane wall. Also important is to mention is that the three-form gauge sector by itself does not determine the detailed dynamics of the charged wall. Once a bubble is present, its wall tension, internal worldvolume structure, and any worldvolume \(U(1)\) monopole flux it supports must be specified by an additional membrane effective theory. Since the object under consideration is a charged brane-like membrane, the natural next step is to model its worldvolume by a Dirac--Born--Infeld action and study the resulting bubble energetics. We now turn to that description.

\section{\centering Mechanics of Vacuum Bubbles}

\subsection{\centering Why the Wall Tension is not determined by the Bulk Three-Form Gauge Sector}
\label{subsec:wall_tension_not_from_bulk}

Vacuum bubbles in an ordinary scalar field theory and vacuum bubbles in a three-form gauge sector are similar only at a coarse level in that in both cases one separates two regions with different vacuum energies. Microscopically, however, the origin of the wall is very different, and this difference is crucial for the present construction.
\\ \\
In a scalar field theory, the wall is not an additional ingredient. Rather, it is a smooth bulk field configuration of the same degree of freedom that also determines the vacuum energies. Consider a scalar field $\phi$ with action,

\begin{equation}
	S_{\phi} = \int d^{4}x \, \left[ -\frac{1}{2} \, \partial_{\mu}\phi \, \partial^{\mu}\phi - V(\phi)
	\right] .
\end{equation}
If the potential has two local minima, $\phi_{\rm f}$ and $\phi_{\rm t}$, then a domain wall or bubble wall is described by a configuration in which $\phi$ interpolates smoothly between these two values. For a locally planar static wall depending only on the transverse coordinate $z$, the energy per unit area is,

\begin{equation}
	\sigma_{\phi} = \int dz \, \left[ \frac{1}{2} \left(\frac{d\phi}{dz}\right)^{2} + V(\phi) - V(\phi_{\rm t})
	\right] .
\end{equation}
Using the first integral of the static field equation in the thin wall regime,

\begin{equation}
	\frac{1}{2} \left(\frac{d\phi}{dz}\right)^{2} = V(\phi) - V(\phi_{\rm t}) ,
\end{equation}
one obtains the familiar expression,

\begin{equation}
	\sigma_{\phi} = \int_{\phi_{\rm t}}^{\phi_{\rm f}} d\phi \, \sqrt{2 \, \bigl(V(\phi) - V(\phi_{\rm t})\bigr)} .
\end{equation}
The pressure driving the bubble is then determined by the bulk vacuum energy
difference,

\begin{equation}
	\Delta V = V(\phi_{\rm f}) - V(\phi_{\rm t}) .
\end{equation}
Thus, in the scalar case, the same bulk Lagrangian determines both the vacuum energies and the wall tension. The wall is a solitonic profile of the bulk field itself, and its tension is obtained by integrating the bulk gradient and potential energy across the interpolation region \cite{ CDL1980, Coleman1977, CallanColeman1977}.
\\ \\
The present setup is structurally different. The bulk gauge variable in the present framework is three-form gauge potential $C_{3}$ with four-form field strength $F_{4} = d C_{3}$. In four spacetime dimensions, the action takes the form,

\begin{equation}
	S_{3} = - \frac{1}{2 \cdot 4!} \int d^{4}x \, \sqrt{-g} \, F_{\mu\nu\rho\sigma} F^{\mu\nu\rho\sigma} + q_{b} \int_{\Sigma_{3}} C_{3} .
\end{equation}
The sourced field equation may be written schematically as,

\begin{equation}
	\nabla_{\mu} F^{\mu\nu\rho\sigma} = J^{\nu\rho\sigma}_{\Sigma} ,
\end{equation}
where $J^{\nu\rho\sigma}_{\Sigma}$ is localized on the membrane worldvolume
$\Sigma_{3}$. Away from the membrane one has $J^{\nu\rho\sigma}_{\Sigma} = 0$, so the bulk equation reduces to (see Appendix B),

\begin{equation}
	\nabla_{\mu} F^{\mu\nu\rho\sigma} = 0 \qquad \Longrightarrow \qquad
	F^{\mu\nu\rho\sigma} = q \, \epsilon^{\mu\nu\rho\sigma} ,
\end{equation}
with $q$ constant on each connected bulk region. In other words, the bulk
solution on each side of the wall is not a smooth finite thickness profile but a constant four-flux branch labelled by $q$.
\\ \\
Integrating the sourced equation across a narrow Gaussian pillbox transverse to the membrane gives the jump condition \eqref{eq:flux_jump},

\begin{equation}
	q_{\rm out} - q_{\rm in} = q_{b} ,
\end{equation}
so the membrane interpolates between two piecewise constant four-flux flux sectors. Hence the bulk sector fixes the vacuum energy difference across the bubble,

\begin{equation}
	\Delta \rho_{\rm vac} = \frac{1}{2} \left( q_{\rm out}^{2} - q_{\rm in}^{2} \right) ,
\end{equation}
but this is precisely where the information supplied by the local bulk theory stops.
\\ \\
The reason is that, unlike the scalar case, there is no smooth bulk interpolation problem whose on-shell energy could be identified with the wall tension. 
\\ \\
The four-form field strength in four spacetime dimensions carries no
local propagating degrees of freedom. Away from charged sources, its
on-shell solution is $F_4 = q \, {\rm vol}_4$, where the constant flux parameter $q$ labels the corresponding vacuum branch. The change from $q_{\rm out}$ to $q_{\rm in}$ is supported by the membrane as a localized object electrically charged under the three-form gauge potential $C_3$. Therefore, there is no analogue here of the scalar formula,

\begin{equation}
	\sigma_{\phi} = \int d\phi \, \sqrt{2 \, (V - V_{\rm t})} ,
\end{equation}
because there is no source-free smooth interpolation of the bulk four-form flux whose integrated energy could determine the membrane tension. Equivalently, the bulk theory determines the vacuum data on the two sides of the wall, but not the microscopic energy stored in the wall.
\\ \\
This is the central structural distinction. In scalar field theory, the wall is emergent from the bulk field profile. In a three-form gauge sector, the wall is an additional charged extended object whose mechanics must be specified by its own worldvolume effective action. At low energies, the most general viewpoint is therefore to treat the membrane tension and its internal degrees of freedom as independent effective parameters,

\begin{equation}
	S_{\rm wall} = - \int_{\Sigma_{3}} d^{3}\xi \, \sqrt{-\gamma} \left[
	T_{\rm eff}	+ \mathcal{L}_{\rm int} \right]	+ q_{b} \int_{\Sigma_{3}} C_{3} ,
\end{equation}
where $\gamma_{ab}$ is the induced metric on the wall, $T_{\rm eff}$ is the effective wall tension, $\xi$ are the wall's worldvolume coordinates and $\mathcal{L}_{\rm int}$ denotes possible worldvolume fields and interactions. The bulk vacuum data $\{ q_{\rm out}, q_{\rm in}, \Delta \rho_{\rm vac} \}$ determine the volume contribution to the bubble energetics, whereas the local mechanics of the wall is controlled by the independent worldvolume data $\{ T_{\rm eff}, \mathcal{L}_{\rm int} \}$.
\\ \\
It is important to stress that this does not mean that a more microscopic
ultraviolet completion can never relate the wall parameters to other scales in the theory. It means only that such relations are not implied by the local bulk three-form gauge sector dynamics alone. In particular, the quantities such as wall tension $T_{\rm eff}$, the worldvolume gauge coupling $g_{\rm YM}$, if present and the detailed nonlinear response of the wall are not fixed by knowing only the branch energies $\Lambda_{\rm eff}(q)$ or the jump condition $q_{\rm out} - q_{\rm in} = q_{b}$. They belong to the membrane sector.
\\ \\
Since the object under consideration is a charged brane-like membrane, the natural minimal nonlinear effective description is the Dirac--Born--Infeld action. This choice is not claimed to be unique, rather, it is adopted as the minimal symmetry respecting worldvolume theory that simultaneously and naturally contains both a tension term and an intrinsic worldvolume $U(1)$ gauge sector. In particular, the DBI form reduces to the usual Maxwell theory at weak field strength while also resumming the nonlinear corrections that become important when the worldvolume
\(U(1)\) magnetic field strength becomes large.
\\ \\
A related distinction also appears in the tunneling problem. In ordinary scalar field theory, vacuum decay is described by an $O(4)$ symmetric Euclidean bounce in the zero-temperature limit, or by an $O(3)$ symmetric bounce when thermal effects dominate \cite{Affleck:1981, Linde:1983}. In the present framework, the analogous non-perturbative event is membrane nucleation between two constant four-form flux branches in the sense of Brown and Teitelboim. A complete computation of the corresponding nucleation exponent would therefore require the coupled Euclidean membrane problem, including gravity, the bulk three-form gauge sector and its four-form flux, together with the membrane worldvolume action.
\\ \\
In the present work, however, our aim is narrower. We do not attempt to derive the nucleation rate from first principles. Rather, we take membrane-mediated branch transitions as the physical origin of the bubbles and focus on the subsequent Lorentzian worldvolume energetics of an already formed bubble. Our primary question is therefore not how frequently such bubbles are nucleated, but what their fate is once they exist. For that question, the essential input is precisely the independent membrane effective theory discussed above.
\\ \\
Furthermore, if an already formed bubble is shown to be stable, with a finite rest energy and no tendency toward runaway expansion or relaxation to a trivial zero-energy state, then it is natural to regard that object as a localized particle-like species rather than merely as a transient instanton configuration. In that case, a Euclidean tunneling analysis is not required in order to establish its existence as a stable remnant. Such an analysis becomes necessary only if one wishes to compute its production rate from a specific nucleation channel, or if one studies the runaway branch for which the bubble expands rather than stabilizes. We will return to this distinction in Secs.~6 and 7.

\subsection{\centering Minimal Worldvolume Effective Theory}

The Dirac-Born-Infeld action reads \cite{PolchinskiBook2, DBI1934, Leigh1989},

\begin{equation}
	S_{b} \; = \; - T_{\rm eff} \int d^{3}\xi \, \sqrt{-\det \bigl(\gamma_{ab} + 2 \pi \alpha' F_{ab}\bigr)} \, ,
	\label{eq:dbi_action_sec5}
\end{equation}
where $T_{\rm eff}$ is the effective wall tension, $\xi^{a}$ are the worldvolume coordinates, $\gamma_{ab}$ is the induced metric on the wall, $\alpha'$ is the Regge slope parameter, and $F_{ab}$ is the field strength of a worldvolume $U(1)$ gauge field living on the membrane. 
\\ \\
For a spherically symmetric wall of instantaneous radius $R$, we work in static worldvolume coordinates and take the induced metric to be,

\begin{equation}
	\gamma_{ab} \; = \; {\rm diag} \bigl(-1, \, R^{2}, \, R^{2} \sin^{2} \theta \bigr) \, .
	\label{eq:induced_metric_sec5}
\end{equation}
This form is sufficient for the present purpose, since our immediate
goal is to determine the radius dependence of the wall energy and to
investigate whether the worldvolume \(U(1)\) flux can convert the
collapsed endpoint from a trivial zero-energy configuration into a
finite-energy remnant.
\\ \\
In the weak field regime, $|2 \pi \alpha' F_{ab}| \ll |\gamma_{ab}|$, the DBI action may be expanded in powers of $F_{ab}$. To quadratic order one finds,

\begin{equation}
	S_{b} \; = \; - T_{\rm eff} \int d^{3}\xi \, \sqrt{-\det \gamma} \; - \; \frac{(2 \pi \alpha')^{2} T_{\rm eff}}{4} \int d^{3} \xi \, \sqrt{-\det \gamma} \, F_{ab} F^{ab} \; + \; \mathcal{O}(F^{4}) \, .
	\label{eq:dbi_expansion_sec5}
\end{equation}
The first term is the usual membrane area term, while the second is the gauge kinetic term induced on the worldvolume. Matching the latter to the canonical Maxwell form in $2 + 1$ dimensions,

\begin{equation}
	S_{\rm kin} \; = \; - \frac{1}{4 g^{2}} \int d^{3}\xi \, \sqrt{-\det \gamma} \, F_{ab} F^{ab} \, ,
	\label{eq:maxwell_kinetic_sec5}
\end{equation}
identifies the Yang--Mills coupling as,

\begin{equation}
	g_{\rm YM}^{2} \; = \; \frac{1}{(2 \pi \alpha')^{2} \, T_{\rm eff}} \, .
	\label{YM_coupling}
\end{equation}
At the level of the low energy membrane effective theory, the quantities
$T_{\rm eff}$, $g_{\rm YM}$, and $\alpha'$ should in general be regarded as
independent matched parameters. For the stability mechanism developed in the
present work, no special relation among them is required. Nevertheless, it is useful to adopt an illustrative ultraviolet motivated benchmark in which these parameters are related in the same way as for a D2-brane in string theory. This is natural because the bubble wall is being modeled as a brane-like membrane carrying a worldvolume $U(1)$ gauge field, and the DBI action of a $D_p$-brane provides the canonical example of such a system. Expanding the DBI action to quadratic order identifies the gauge kinetic term and hence relates $g_{\rm YM}$ to $\alpha'$ and the brane tension, while the standard $D_p$-brane tension formula relates $\alpha'$ and the tension itself. Combining the two then yields a string-inspired relation between $g_{\rm YM}$ and $T_{\rm eff}$. In the present work we use this relation only as a convenient benchmark for numerical estimates and dimensional normalization. The existence of the collapsed flux-supported remnant does not depend on this particular stringy identification. The standard string theoretic expression for the tension of a $D_p$ brane \cite{PolchinskiBook1} reads,

\begin{equation}
	T_{p} \; = \; \frac{1}{(2 \pi)^{p} g_{s}} \, \alpha'^{-\frac{p + 1}{2}} \, .
	\label{eq:Tp_general_sec5}
\end{equation}
Since the bubble wall is a two dimensional spatial membrane, we set $p = 2$, which gives,

\begin{equation}
	T_{2} \; = \; \frac{1}{(2 \pi)^{2} g_{s}} \, \alpha'^{- \frac{3}{2}} \, .
\end{equation}
Solving for $\alpha'$ then yields,

\begin{equation}
	\alpha' \; = \; (2 \pi)^{-4/3} \, g_{s}^{-2/3} \, T_{2}^{-2/3} \, .
	\label{eq:alpha_prime_sec5}
\end{equation}
Upon substituting this into \eqref{YM_coupling}, and identifying $T_{2}$ with the effective wall tension $T_{\rm eff}$ relevant for the present membrane, one obtains,

\begin{equation}
	g_{\rm YM}^{2} \; = \; (2 \pi)^{2/3} \, g_{s}^{4/3} \, T_{\rm eff}^{1/3} \, .
	\label{eq:gym_general}
\end{equation}
For an illustrative benchmark choice, say $g_{s} = 1 / \sqrt{2 \pi}$, this simplifies to the particularly transparent relation,

\begin{equation}
	g_{\rm YM}^{2} \; = \; T_{\rm eff}^{1/3} \qquad \Longrightarrow \qquad
	g_{\rm YM} \; = \; T_{\rm eff}^{1/6} \, .
	\label{eq:gym_benchmark}
\end{equation}
This benchmark is adopted only for convenience, since it allows the gauge coupling and the wall tension to be related by a simple power law. The string coupling $g_{s}$ should otherwise be regarded as a free parameter subject only to the perturbative condition $0 < g_{s} < 1$. Our later numerical estimates make use of the benchmark choice above, but the qualitative conclusions do not depend on this particular normalization.
\\ \\
In principle, the worldvolume gauge coupling may acquire scale dependence through quantum corrections. However, in $2+1$ dimensions the coupling $g_{\rm YM}^{2}$ is dimensionful, with mass dimension one, so the usual notion of logarithmic running must be treated with care. In the minimal approximation where the membrane supports only a pure abelian worldvolume $U(1)$ gauge sector, that is, with no self interaction, and assuming that the worldvolume does not host any charged matter fields or any higher derivative operators (all of which are not part of the effective DBI action), one may regard $g_{\rm YM}^{2}$ as a matched effective parameter at some reference scale $\mu_{0}$,

\begin{equation}
	g_{\rm YM}^{2}(\mu_{0}) \; = \; \frac{1}{(2 \pi \alpha')^{2} \, T_{\rm eff}} \, ,
\end{equation}
with no significant perturbative running of the dimensionful coupling itself over the regime of validity of the effective theory. The corresponding dimensionless interaction strength is then,

\begin{equation}
	\hat{g}^{2}(\mu) \; \equiv \; \frac{g_{\rm YM}^{2}(\mu)}{\mu} \, ,
\end{equation}
which, in the absence of additional quantum corrections, obeys the classical scaling law,

\begin{equation}
	\mu \frac{d \hat{g}^{2}}{d\mu} \; = \; - \hat{g}^{2} \, .
\end{equation}
With this in mind, we note that the role of the worldvolume gauge sector becomes immediately clear if one first considers the trivial case in which no worldvolume \(U(1)\) flux is supported on the membrane. Setting $F_{ab} = 0$ in \eqref{eq:dbi_action_sec5}, the DBI action reduces to the pure area term,

\begin{equation}
	S_{b} \; = \; - T_{\rm eff} \int d\xi^{0} \int d^{2}\xi \, \sqrt{\det \gamma_{ij}} \; = \; - 4 \pi \, T_{\rm eff} \, R^{2} \int d\xi^{0} \, .
	\label{eq:tension_only}
\end{equation}
The corresponding rest energy is therefore,

\begin{equation}
	E_{\rm wall}(R) \; \sim \; 4 \pi \, T_{\rm eff} \, R^{2} \, ,
	\label{eq:tension_energy_sec5}
\end{equation}
which decreases monotonically as the radius shrinks and vanishes in the limit $R \to 0$. In other words, tension alone always favors collapse and cannot by itself support a pointlike or collapsed relic with finite mass.
\\ \\
This observation is central to the logic of the present work. If the bubble wall carried no internal worldvolume structure, then a collapsing bubble would simply relax to zero size and zero energy. The existence of a stable finite-energy remnant therefore requires an additional conserved ingredient capable of preventing the collapsed endpoint from becoming a trivial zero-energy configuration. In the present framework, that ingredient is the conserved monopole flux carried by the worldvolume gauge field. The nonlinear DBI dynamics then determine whether the collapsing bubble
can approach a stable endpoint with finite nonzero rest energy. We now turn to that question.

\subsection{\centering Topological Worldvolume $U(1)$ Flux on the Bubble Wall}

\begin{figure}[htbp] 
	\centering 
	\includegraphics[width = \textwidth]{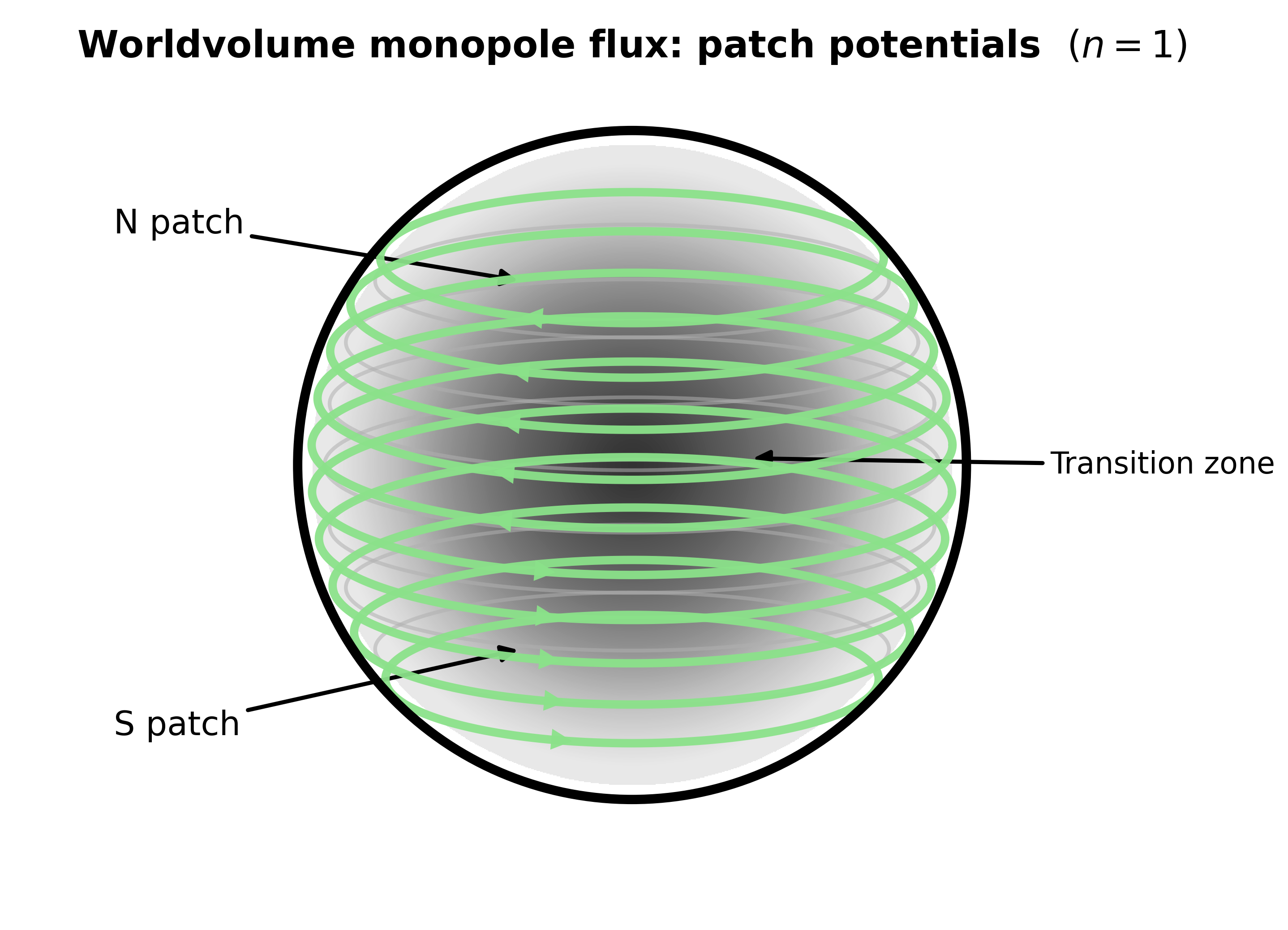} \caption{Patchwise description of the worldvolume gauge potential for a monopole flux on the bubble ($n = 1$). The light-green azimuthal circulation indicates the local form of the potential $A$ in the north and south patches (clockwise in the north patch; anticlockwise in the south patch), which are related on the overlap by a gauge transformation, which can be written schematically as $A_S = A_N - n \, d \phi$. The physical magnetic flux is the globally defined 2-form $F = dA$ with quantized total flux through the bubble.} \label{n = 1 Topolon}
\end{figure}

A nontrivial collapsed configuration requires more than the tension term alone. As shown in the previous subsection, if the worldvolume gauge sector is trivial, the wall energy decreases monotonically as the bubble shrinks and vanishes in the limit $R \to 0$. To obtain a collapsed object with nonzero rest energy, the worldvolume $U(1)$ gauge field must therefore carry a conserved topological charge. The simplest possibility is a Dirac-monopole-type magnetic flux on the spherical wall \cite{WuYang1975,Nakahara2003}. Such a configuration is topologically nontrivial and is characterized by a nonvanishing first Chern class, $c_{1}(L) \neq 0$, for the associated complex line bundle over $S^{2}$ \cite{Nakahara2003, Chern1946, MilnorStasheff1974}.
\\ \\
Because a monopole gauge potential cannot be defined globally on the two sphere, the gauge field must be described using at least two overlapping coordinate patches. We therefore introduce a north patch $(N)$ and a south patch $(S)$, with gauge potentials,

\begin{equation}
	A^{(N)}_{\phi} \; = \; \frac{n}{2} \, \bigl(1 - \cos\theta\bigr) \, ,
	\qquad
	A^{(S)}_{\phi} \; = \; - \frac{n}{2} \, \bigl(1 + \cos\theta\bigr) \, ,
	\label{eq:monopole_potential}
\end{equation}
which are regular away from the south and north poles respectively. On the overlap of the two patches, these gauge potentials differ by a gauge transformation with transition function $\exp(i n \phi)$. Single-valuedness of the bundle then requires $n \in \mathbb{Z}$, so the monopole number is quantized. Equivalently, the integer $n$ is the first Chern number, or the total magnetic flux in units of $2 \pi$, \footnote{It is important to distinguish the two charges appearing in the construction. The quantity $q_b$ is the membrane's electric charge under the bulk three-form gauge potential $C_3$ and controls the jump in the bulk four-form flux. By contrast, $n \in \mathbb{Z}$ is the first Chern number of the intrinsic worldvolume $U(1)$ gauge bundle and quantizes the magnetic flux-supported on the membrane. These are distinct data of the effective theory.}

\begin{equation}
	n \; = \; c_{1}(L) \; = \; \frac{1}{2 \pi} \int_{S^{2}} F \;\in\; \mathbb{Z} \, .
	\label{eq:chern_number}
\end{equation}
Thus $n$ labels disconnected topological sectors of the worldvolume gauge field, and cannot change continuously under smooth deformations of the configuration.
\\ \\
For the patchwise potentials in \eqref{eq:monopole_potential}, the field strength is globally well defined and is given by
\begin{equation}
	F_{\theta\phi} \; = \; \partial_{\theta} A_{\phi} \; = \; \frac{n}{2} \, \sin\theta \, .
	\label{eq:F_thetaphi_monopole}
\end{equation}
The total worldvolume magnetic flux is therefore fixed entirely by the
integer $n$, independently of the bubble radius. This is the key point physically. As the wall contracts, this conserved worldvolume flux cannot change continuously and is compressed into a progressively smaller physical
area. The resulting growth of the gauge field energy is what makes a flux supported remnant possible.
\\ \\
To see this explicitly, consider first the Maxwell limit of the DBI action. For a spherical wall of radius $R$, the magnetic contribution to the gauge field energy is,

\begin{equation}
	E_{G}(R) \; = \; \frac{1}{4 g_{\rm YM}^{2}} \int_{S^{2}} d^{2}\xi \, \sqrt{\det \gamma_{ij}} \, F_{ab} F^{ab} \; = \; \frac{1}{4 g_{\rm YM}^{2}} \int_{S^{2}} d^{2}\xi \, \sqrt{\det \gamma_{ij}} \, 2 \, F_{\theta\phi} F^{\theta\phi} \, .
	\label{eq:monopole_energy_maxwell}
\end{equation}
Substituting \eqref{eq:F_thetaphi_monopole} in the equation above gives,

\begin{equation}
	E_{G}(R) \; = \; \frac{n^{2} \pi}{2 g_{\rm YM}^{2}} \, \frac{1}{R^{2}} \, .
\end{equation}
Using the benchmark relation \eqref{eq:gym_benchmark}, this may be written as,

\begin{equation}
	E_{G}(R) \; = \; \frac{n^{2} \pi}{2 T_{\rm eff}^{1/3}} \, \frac{1}{R^{2}} \, .
	\label{eq:monopole_energy_maxwell_benchmark}
\end{equation}
The $R^{-2}$ scaling is the central result. For fixed monopole number $n$, shrinking the bubble forces the same quantized flux through a smaller and smaller area, so the magnetic energy grows rapidly as $R$ decreases. In the pure Maxwell truncation, this contribution diverges in the limit $R \to 0$.
\\ \\
This divergence should not be interpreted as a pathology of the physical configuration itself, but rather as a signal that the Maxwell approximation becomes inadequate in the strongly compressed regime. The full DBI action is precisely designed to capture this regime by resumming the higher powers of $F_{ab}$ that are neglected in the quadratic truncation. As we shall show in Sec.~6, once the complete DBI structure is retained, the wall energy no longer diverges without bound in the collapsed limit. Instead, the nonlinear worldvolume dynamics regulate the short distance behavior and yield a finite nonzero rest energy for the collapsed flux carrying object. It is this finite energy pointlike remnant that we identify as a topolon. Bachas et al.~\cite{Bachas:2000} studied the stabilization of an
existing extended D-brane against shrinking by a quantized worldvolume \(U(1)\) flux. By contrast, the present mechanism concerns a branch-changing vacuum bubble in four dimensions. Here the membrane separates two constant four-form flux branches, and the worldvolume flux does not merely stabilize an extended brane at finite size, but can drive the collapse toward a finite-mass localized remnant rather than either runaway expansion or relaxation to a trivial zero-energy state.
\\ \\
The corresponding north--south patch construction, together with the gauge transition on the overlap, is illustrated in Fig.\ref{n = 1 Topolon} for the case $n = 1$.

\FloatBarrier
\noindent Having identified the relevant bulk four-form flux-jump channel and the worldvolume $U(1)$ monopole flux responsible for the nontrivial
collapsed endpoint, we now combine the vacuum-volume contribution with the full DBI wall energy.

\section{\centering Collapsed Vacuum Bubbles Carrying Non-Zero Worldvolume Monopole Flux}

\subsection{\centering The Bubble Energy in the Full DBI Theory}

We now quantify the energetics of a vacuum bubble carrying quantized
worldvolume $U(1)$ monopole flux within the full $2 + 1$-dimensional
Dirac--Born--Infeld description of the membrane wall.

\begin{equation}
	S \; = \; - T_{\rm eff} \int d^{3}\xi \, \sqrt{-\det\!\bigl(\gamma_{ab} + 2 \pi \alpha' F_{ab}\bigr)} \, ,
	\label{eq:sec6_dbi_action}
\end{equation}
and we take the induced worldvolume metric to be,

\begin{equation}
	\gamma_{ab} \; = \; {\rm diag}\!\bigl(-1, \, R^{2}, \, R^{2}\sin^{2}\theta \bigr) \, .
	\label{eq:sec6_metric}
\end{equation}
We endow the worldvolume $U(1)$ with a quantized monopole flux, as discussed in Sec.~5.3, so that,

\begin{equation}
	\int_{S^{2}} F \; = \; 2 \pi n \qquad \Longrightarrow \qquad
	F_{\theta\phi} \; = \; \frac{n}{2} \sin\theta \, ,
	\label{eq:sec6_flux}
\end{equation}
and we restrict to the static magnetic configuration, for which $F_{t\theta} = F_{t\phi} = 0$. With \eqref{eq:sec6_metric} and \eqref{eq:sec6_flux}, the determinant in \eqref{eq:sec6_dbi_action} factorizes into the time direction and the spatial $2 \times 2$ block,

\begin{equation}
	\det \bigl(\gamma_{ab} + 2 \pi \alpha' F_{ab}\bigr)
	\; = \; \gamma_{tt} \, \det
	\begin{pmatrix}
		R^{2} & 2 \pi \alpha' F_{\theta\phi} \\
		- 2 \pi \alpha' F_{\theta\phi} & R^{2}\sin^{2}\theta
	\end{pmatrix} \, .
	\label{eq:sec6_det_factor}
\end{equation}
Evaluating the $2 \times 2$ determinant and using \eqref{eq:sec6_flux} gives,

\begin{equation}
	- \det \bigl(\gamma_{ab} + 2 \pi \alpha' F_{ab}\bigr) \; = \; R^{4}\sin^{2}\theta + (2 \pi \alpha')^{2} F_{\theta\phi}^{2} \; = \; \sin^{2}\theta \left( R^{4} + (2 \pi \alpha')^{2} \frac{n^{2}}{4} \right) .
	\label{determinant_value}
\end{equation}
The static wall energy is obtained by integrating the DBI energy density over $S^{2}$,

\begin{equation}
	E_{bw}^{(\rm DBI)}(R) \; = \; T_{\rm eff} \int_{S^{2}} d\theta \, d\phi \, \sqrt{-\det \bigl(\gamma_{ab} + 2 \pi \alpha' F_{ab}\bigr)} \, ,
\end{equation}
so using \eqref{determinant_value} and

\begin{equation}
	\int_{S^{2}} d\theta \, d\phi \, \sin\theta \; = \; 4 \pi
\end{equation}
yields,

\begin{equation}
	E_{bw}(R) \; = \; 4 \pi \, T_{\rm eff} \, \sqrt{R^{4} + (2 \pi \alpha')^{2} \, \frac{n^{2}}{4}} \, .
	\label{eq:sec6_Ebw_alpha}
\end{equation}
Using the relation \eqref{YM_coupling}, we may rewrite \eqref{eq:sec6_Ebw_alpha} as,

\begin{equation}
	E_{bw}(R) \; = \; 4 \pi \, T_{\rm eff} \, \sqrt{R^{4} + \frac{n^{2}}{4 g_{\rm YM}^{2} \, T_{\rm eff}}} \, .
	\label{eq:BubbleWallEnergy}
\end{equation}
This is the exact DBI wall energy, within the static spherical ansatz,
for a membrane carrying worldvolume monopole number $n$. In contrast to the Maxwell approximation discussed in Sec.~5.3, the full DBI expression remains finite as $R \to 0$. Within the DBI effective description, the nonlinear worldvolume dynamics render the wall energy finite in the formal collapsed limit $R \to 0$, thereby allowing a finite-energy remnant rather than a trivial zero-energy endpoint.
\\ \\
The total bubble energy is then,

\begin{equation}
	E_{t}(R) \; = \; E_{bw}(R) + E_{\rm vac}(R) \; = \; 4 \pi \, T_{\rm eff} \, \sqrt{R^{4} + \frac{n^{2}}{4 g_{\rm YM}^{2} \, T_{\rm eff}}}
	\; + \; \frac{4 \pi}{3} R^{3} \, \Delta \rho \, ,
	\label{eq:sec6_total_energy}
\end{equation}
where the second term is the volume contribution arising from the
vacuum-energy difference between the interior and exterior bulk
four-form flux branches. As discussed in Sec.~4, the class of admissible branch transitions studied in this work satisfies $\Delta \rho \geq 0$, with the exact sign-flip channel giving $\Delta \rho = 0$ and more generic admissible jumps giving $\Delta \rho > 0$. Nevertheless, it is useful to analyze the three possible signs of $\Delta \rho$ separately, since this makes the stability criterion completely transparent.

\subsection{\centering Small Radius Expansion and the Collapsed Limit}

\begin{figure}[htbp] 
	\centering 
	\includegraphics[width = \textwidth]{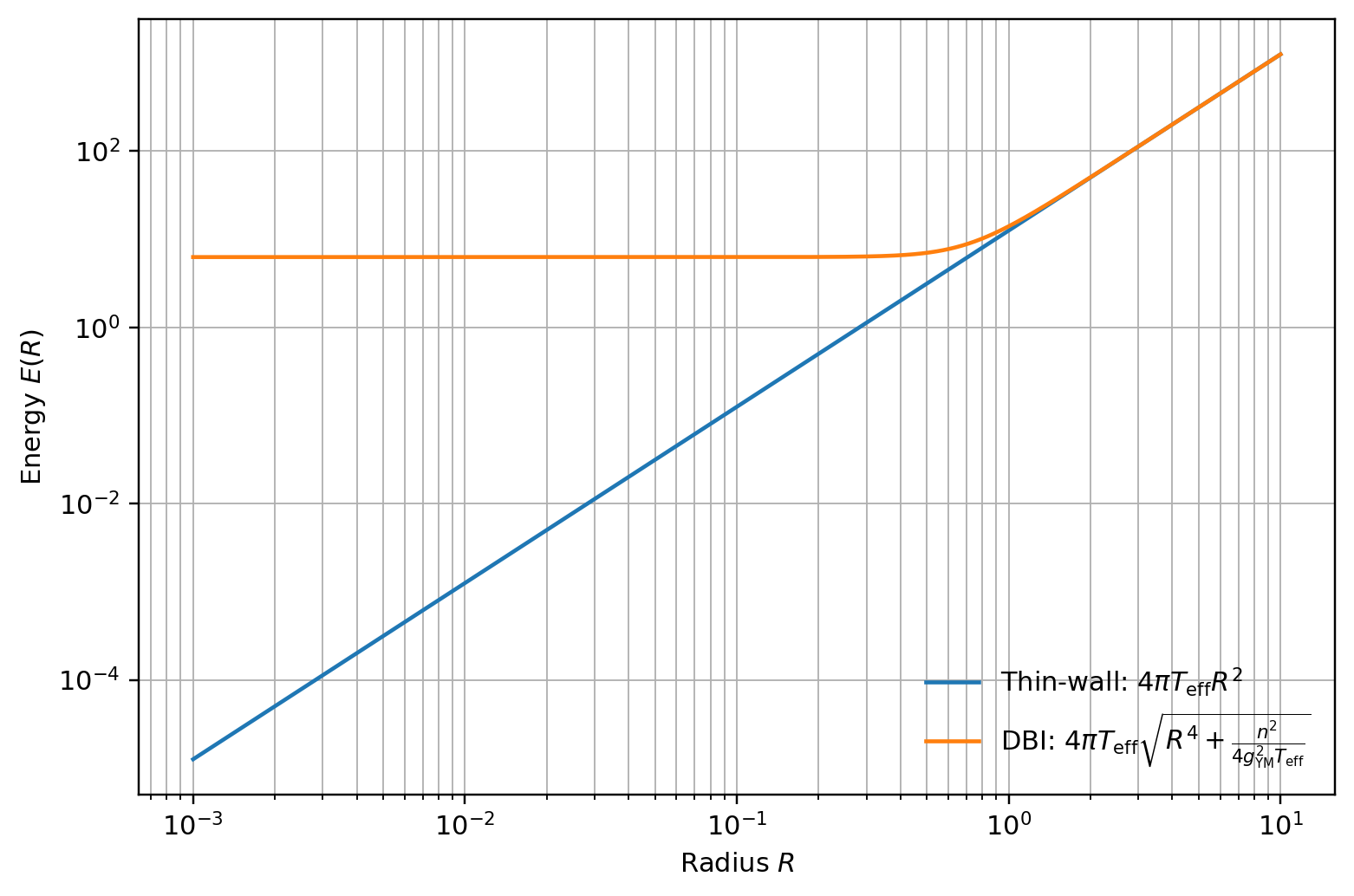} \caption{\footnotesize{Comparison of the thin-wall energy and the full DBI energy of a flux-carrying bubble as a function of radius $R$. Both axes are shown on logarithmic scales. The thin-wall contribution $E_{\rm wall}(R) = 4 \, \pi T_{\rm eff} \, R^2$ vanishes as $R\to 0$, whereas the DBI completion $E_{\rm DBI}(R) = 4 \, \pi \, T_{\rm eff} \, \sqrt{R^4 + \dfrac{n^2}{4 \, g_{\rm YM}^2 \, T_{\rm eff}}}$ approaches a finite constant set by the quantized worldvlume $U(1)$ flux. The plot uses dimensionless, order-one parameter choices and arbitrary overall units. Its purpose is purely illustrative and is to highlight the different small-$R$ behavior of the two energy functions.} } 
	\label{Full DBI vs Maxwell Comparison} 
\end{figure}

To streamline the analysis, let us define,

\begin{equation}
	\mathcal{A} \; \equiv \; \frac{n^{2}}{4 g_{\rm YM}^{2} \, T_{\rm eff}} \, .
	\label{eq:A_def}
\end{equation}
Then the total energy \eqref{eq:sec6_total_energy} takes the compact form,

\begin{equation}
	E_{t}(R) \; = \; 4 \pi \, T_{\rm eff} \, \sqrt{R^{4} + \mathcal{A}}
	\; + \; \frac{4 \pi}{3} R^{3} \, \Delta \rho \, .
	\label{eq:Et_compact}
\end{equation}
The collapsed limit is now immediate. Setting $R = 0$ gives,

\begin{equation}
	E_{t}(0) \; = \; 4 \pi \, T_{\rm eff} \, \sqrt{\mathcal{A}}
	\; = \; \frac{2 \pi \, |n| \, \sqrt{T_{\rm eff}}}{g_{\rm YM}} \, .
	\label{eq:collapsed_mass_general}
\end{equation}
Thus, for nonzero monopole number $n$, the bubble does not relax to zero energy in the collapsed limit. Instead, it approaches a finite rest energy determined by the membrane parameters and the conserved worldvolume monopole number. For the benchmark relation \eqref{eq:gym_benchmark}, this becomes,

\begin{equation}
	E_{t}(0) \; = \; 2 \pi \, |n| \, T_{\rm eff}^{1/3} \, .
	\label{eq:collapsed_mass_benchmark}
\end{equation}
This is the mass of the collapsed flux carrying remnant. Defining the fixed mass scale

\begin{equation}
	\kappa \; \equiv \; 2 \, \pi \, T_{\rm eff}^{1/3} \, ,
\end{equation}
the collapsed-state spectrum may be written as,

\begin{equation}
	E_t (0) \equiv E_n = \kappa \, |n| \, .
	\label{En definition}
\end{equation}
Thus, within this benchmark normalization, the remnant mass grows
linearly with the absolute value of the worldvolume monopole number.
\\ \\
A comparison between the thin wall energy $E_{\rm wall}(R) = 4\pi T_{\rm eff}R^{2}$, which vanishes as $R \to 0$, and the DBI energy \eqref{eq:BubbleWallEnergy}, which approaches the finite constant \eqref{eq:collapsed_mass_general}, is shown in Fig.\ref{Full DBI vs Maxwell Comparison}. The plot uses arbitrary dimensionless parameter choices and is intended only to illustrate the distinct small $R$ behavior.

\FloatBarrier
\noindent To determine whether the collapsed configuration is stable, it is useful to expand \eqref{eq:Et_compact} at small $R$. Using
\begin{equation}
	E_{t}(R) \; = \; 4 \pi \, T_{\rm eff} \, \sqrt{R^{4} + \mathcal{A}}
	\; + \; \frac{4 \pi}{3} \, \Delta \rho \, R^{3} \, ,
\end{equation}
we factor out $\mathcal{A}$ from the square root to obtain,

\begin{equation}
	\sqrt{R^{4} + \mathcal{A}} \; = \; \sqrt{\mathcal{A}} \, \sqrt{1 + \frac{R^{4}}{\mathcal{A}}} \, .
\end{equation}
For $R^{4} \ll \mathcal{A}$, we may use the binomial expansion,

\begin{equation}
	\sqrt{1 + x} \; = \; 1 + \frac{x}{2} + \mathcal{O}(x^{2}) \, .
\end{equation}
With $x = R^{4}/\mathcal{A}$, we have,

\begin{equation}
	\sqrt{R^{4} + \mathcal{A}} \; = \; \sqrt{\mathcal{A}} \left( 1 + \frac{R^{4}}{2 \mathcal{A}} + \mathcal{O}(R^{8}) \right) \; = \; \sqrt{\mathcal{A}} \; + \; \frac{R^{4}}{2 \sqrt{\mathcal{A}}} \; + \; \mathcal{O}(R^{8}) \, .
\end{equation}
Substituting this into $E_{t}(R)$ yields,

\begin{align}
	E_{t}(R) & \; = \; 4 \pi \, T_{\rm eff} \left( \sqrt{\mathcal{A}} + \frac{R^{4}}{2 \sqrt{\mathcal{A}}} + \mathcal{O}(R^{8}) \right)	+ \frac{4 \pi}{3} \, \Delta \rho \, R^{3} \nonumber \\
	& \; = \; 4 \pi \, T_{\rm eff} \, \sqrt{\mathcal{A}} \; + \; \frac{4 \pi}{3} \, \Delta \rho \, R^{3}	\; + \; \frac{2 \pi \, T_{\rm eff}}{\sqrt{\mathcal{A}}} \, R^{4} \; + \; \mathcal{O}(R^{8}) \, .
\end{align}
Using \eqref{eq:collapsed_mass_general} in the equation above we finally obtain,

\begin{equation}
	E_{t}(R) \; = \; E_{t}(0) \; + \; \frac{4 \pi}{3} \, \Delta \rho \, R^{3} \; + \; \frac{2 \pi \, T_{\rm eff}}{\sqrt{\mathcal{A}}} \, R^{4}
	\; + \; \mathcal{O}(R^{8}) \, .
	\label{small R expansion}
\end{equation}
This expansion shows that the sign of $\Delta \rho$ controls the leading deformation away from the collapsed state. If $\Delta \rho \neq 0$, the cubic volume term dominates the quartic DBI correction at sufficiently small radius. If $\Delta \rho = 0$, the leading correction is positive and quartic.
\\ \\
It is also useful to compute the first derivative of the full energy,
\begin{equation}
	\frac{dE_{t}}{dR} \; = \; \frac{8 \pi \, T_{\rm eff} \, R^{3}}{\sqrt{R^{4} + \mathcal{A}}} \; + \; 4 \pi \, \Delta \rho \, R^{2}
	\; = \; 4 \pi \, R^{2} \left[ \frac{2 T_{\rm eff} \, R}{\sqrt{R^{4} + \mathcal{A}}} + \Delta \rho	\right] .
	\label{eq:Et_derivative}
\end{equation}
This form makes the global behavior of the energy especially clear.

\subsection{\centering The Case $\Delta \rho = 0$}

We begin with the \textit{exact sign-flip channel} discussed in Sec.~4, for which the interior and exterior vacuum energies are degenerate and therefore,

\begin{equation}
	\Delta \rho \; = \; 0 \, .
\end{equation}
In this case the total energy reduces to the DBI wall contribution alone,

\begin{equation}
	E_{t}(R) \; = \; 4 \pi \, T_{\rm eff} \, \sqrt{R^{4} + \mathcal{A}} \, .
	\label{eq:Et_Delta0}
\end{equation}
Its derivative is,

\begin{equation}
	\frac{dE_{t}}{dR} \; = \; \frac{8 \pi \, T_{\rm eff} \, R^{3}}{\sqrt{R^{4} + \mathcal{A}}} \, ,
\end{equation}
which is strictly positive for every $R > 0$ and vanishes only at $R = 0$. Hence $E_{t}(R)$ is a monotonically increasing function of the radius. The energetically preferred configuration is therefore the collapsed limit,

\begin{equation}
	R \; \to \; 0 \, ,
\end{equation}
and the bubble relaxes to the finite mass state \eqref{eq:collapsed_mass_general}, or the benchmark case \eqref{eq:collapsed_mass_benchmark}. In other words, when the volume term is absent, the DBI wall dynamics together with the conserved monopole flux lead directly to a stable collapsed remnant.
\\ \\
To determine the nature of the minimum more explicitly, we substitute $\Delta \rho = 0$ into the small radius expansion \eqref{small R expansion}, which gives,

\begin{equation}
	E_{t}(R) \; = \; E_{t}(0) \; + \; \frac{2 \pi \, T_{\rm eff}}{\sqrt{\mathcal{A}}} \, R^{4} \; + \; \mathcal{O}(R^{8}) \, .
\end{equation}
Thus the leading correction about the collapsed configuration is positive and quartic in $R$. It follows that $R = 0$ is a stable boundary minimum of the static spherical energy functional within the DBI effective description. In particular, the absence of a quadratic or cubic destabilizing term means that small radial deformations away from the collapsed state are energetically suppressed, with the first nonvanishing positive energetic correction appearing at quartic order.
\\ \\
This case is the cleanest one conceptually, since the entire remnant mass comes from the wall tension and the topological worldvolume flux. There is no residual vacuum pressure trying either to expand or to contract the bubble. The exact sign-flip transition is therefore the simplest realization of a topolon.

\subsection{\centering The Case $\Delta \rho > 0$}

We next consider the more general transitions between HHW-admissible four-form flux branches, for which,

\begin{equation}
	\Delta \rho \; > \; 0 \, .
\end{equation}
As already shown in Sec.~4, these are precisely the transitions for which the interior branch remains inside the admissible interval and the vacuum-energy difference is non-negative. The exact sign-flip channel is then recovered only as the limiting case $\Delta \rho = 0$.
\\ \\
For $\Delta \rho > 0$, equation \eqref{eq:Et_derivative} immediately gives,

\begin{equation}
	\frac{dE_{t}}{dR} \; = \; 4 \pi \, R^{2} \left[ \frac{2 T_{\rm eff} \, R}{\sqrt{R^{4} + \mathcal{A}}} + \Delta \rho \right] \; > \; 0 \qquad \text{for all} \qquad R > 0 \, .
	\label{eq:Et_derivative_positive}
\end{equation}
Thus the total energy is again monotonically increasing with radius. The bubble therefore lowers its energy by shrinking, and the global minimum remains at,

\begin{equation}
	R \; \to \; 0 \, .
\end{equation}
The positive vacuum term does not destabilize the remnant. Rather, it makes larger radii even more costly, thereby strengthening the tendency toward collapse. The endpoint of the evolution is again a finite-energy-flux supported remnant with mass \eqref{eq:collapsed_mass_general} or equivalently \eqref{eq:collapsed_mass_benchmark}.
\\ \\
The stability of the collapsed configuration may also be seen directly from the small radius expansion. For $\Delta \rho > 0$, equation \eqref{small R expansion} becomes,
\begin{equation}
	E_{t}(R) \; = \; E_{t}(0) \; + \; \frac{4 \pi}{3} \, \Delta \rho \, R^{3} \; + \; \frac{2 \pi \, T_{\rm eff}}{\sqrt{\mathcal{A}}} \, R^{4}
	\; + \; \mathcal{O}(R^{8}) \, .
\end{equation}
The leading deformation away from $R = 0$ for sufficiently small $R$ is therefore positive and cubic, with the quartic DBI term providing an additional positive correction. Hence the energy increases for sufficiently small positive $R$, confirming that the collapsed configuration is a stable minimum. Relative to the special case $\Delta \rho = 0$, the presence of a positive vacuum contribution strengthens the stability of the remnant.
\\ \\
This result is important. It shows that the exact sign-flip channel is not unique in yielding a stable collapsed object. It is merely the cleanest channel because it removes the volume term altogether. More general admissible branch changes with $\Delta \rho > 0$ also collapse and remain stable once the full DBI wall energy and the conserved monopole flux are taken into account.
\\ \\
To illustrate the qualitative behavior of the bubble energy, we adopt the benchmark parameter choices summarized in Table~\ref{tab:benchmark_energy_plot} below.

\begin{figure}[htbp]
	\centering
	\includegraphics[width=\textwidth]{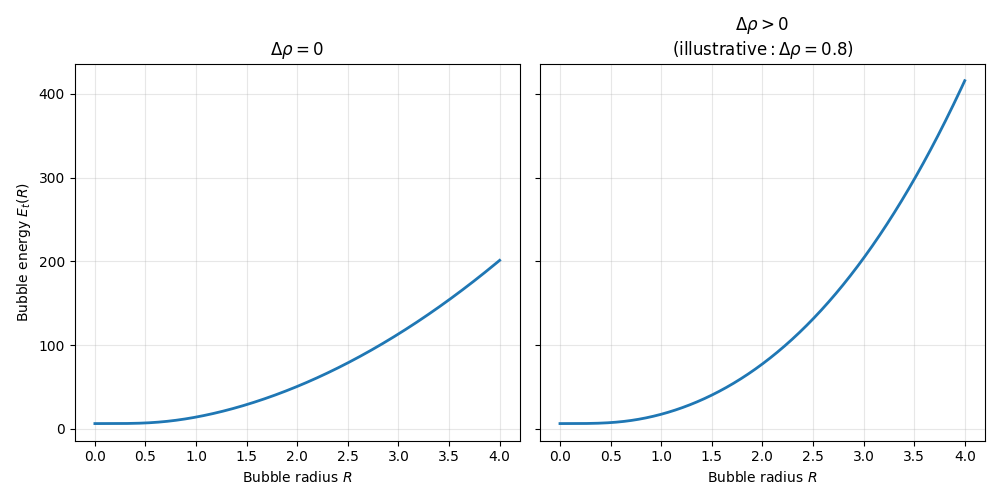}
	\caption{\footnotesize{Illustrative behavior of the total bubble energy $E_{t}(R)$ as a function of the bubble radius $R$ for the two physically admissible cases with $\Delta \rho \geq 0$. Panel (a) shows the case $\Delta \rho = 0$, for which the energy has a stable minimum at $R = 0$ and rises quartically about the collapsed configuration. Panel (b) shows the case $\Delta \rho > 0$, for which the collapsed state remains the global minimum and the additional positive cubic term further strengthens stability. Note that in both cases, the energy of the bubble at $R = 0$ is not zero but is finite and is equal to $2 \, \pi$. The plots are generated using the benchmark values listed in Table~\ref{tab:benchmark_energy_plot}, and are intended only to display the qualitative dependence of the energy profile on the sign of $\Delta \rho$.}}
	\label{fig:bubble_energy_positive_cases}
\end{figure}

\begin{table}[htbp]
	\centering
	\caption{\footnotesize{Illustrative benchmark values used in the energy plots of Figs.~\ref{fig:bubble_energy_positive_cases}, \ref{fig:bubble_energy_negative_case}, \ref{fig:bubble_energy_negative_case_2} and \ref{fig:bubble_energy_negative_case_3}.}}
	\label{tab:benchmark_energy_plot}
	\begin{tabular}{ccc}
		\hline
		Parameter & Value & Comment \\
		\hline
		$T_{\rm eff}$ & $1$ & Effective wall tension \\
		$g_{\rm YM}$ & $1$ & Worldvolume gauge coupling \\
		$n$ & $1$ & Monopole flux number \\
		$\mathcal{A} = \dfrac{n^{2}}{4 g_{\rm YM}^{2} \, T_{\rm eff}}$ & $\dfrac{1}{4}$ & Derived combination entering $E_{t}(R)$ \\
		\hline
	\end{tabular}
\end{table}
\FloatBarrier
\noindent For these values, one has,

\begin{equation}
	\mathcal{A} \; = \; \frac{n^{2}}{4 g_{\rm YM}^{2} \, T_{\rm eff}} \; = \; \frac{1}{4}
\end{equation}
and the energy of finite-mass remnant, given by equation \eqref{eq:collapsed_mass_general}, then reads,

\begin{equation}
	E_t(0) = 4 \pi \, T_{\rm eff} \sqrt{\mathcal{A}} = 2 \, \pi \approx 6.28 \, .
\end{equation}
These benchmark values are chosen solely for visualization. They are not meant to represent a unique physical normalization, but only to provide a simple reference point for plotting the energy profile. The stability conclusions derived above do not depend on this particular choice, and instead rely only on the conditions $T_{\rm eff} > 0$, $g_{\rm YM}^{2} > 0$, and $\mathcal{A} > 0$.
\\ \\
 The resulting energy profiles for the two physically admissible cases, $\Delta \rho = 0$ and $\Delta \rho > 0$, are shown in Fig.~\ref{fig:bubble_energy_positive_cases}. In both cases, the energy is minimized at $R = 0$, confirming that the bubble collapses to a finite-energy remnant rather than dispersing or undergoing runaway expansion. The case $\Delta \rho = 0$ exhibits quartic stabilization about the collapsed state, whereas for $\Delta \rho > 0$ the positive cubic contribution makes the growth of the energy away from $R = 0$ even more pronounced.

\FloatBarrier

\subsection{\centering The Case $\Delta \rho < 0$}

\begin{figure}[htbp]
	\centering
	\includegraphics[width=0.65\textwidth]{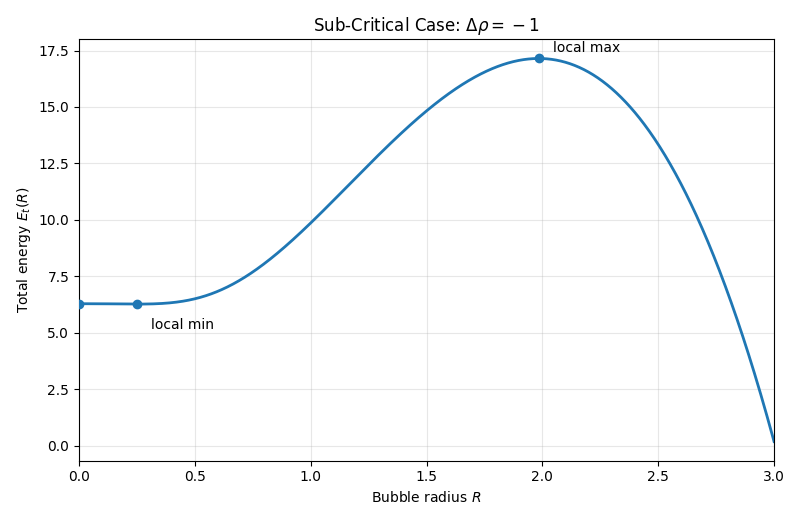}
	\caption{\footnotesize{Illustrative behavior of the total bubble energy $E_{t}(R)$ for the contrast case $\Delta \rho < 0$, generated using the same benchmark values as in Table~\ref{tab:benchmark_energy_plot}. For the illustrative choice $\Delta \rho = -1$, we are in the sub-critical regime and therefore there is a local minimum, followed by a local maximum in the energy profile. Indeed, we note that the energy initially decreases away from the collapsed configuration at $R = 0$ and develops a local minimum at $R \approx 0.252$, corresponding to a metastable finite-radius bubble. This is then followed by a local maximum at $R \approx 1.984$, corresponding to the barrier beyond which the negative vacuum term drives runaway expansion. The energy of the bubble at $R = 0$, $R \approx 0.252$ and $R \approx 1.984$ is $6.28$, $6.266$ and $17.149$ respectively.}}
	\label{fig:bubble_energy_negative_case}
\end{figure}

\begin{figure}[htbp]
	\centering
	\includegraphics[width=0.65\textwidth]{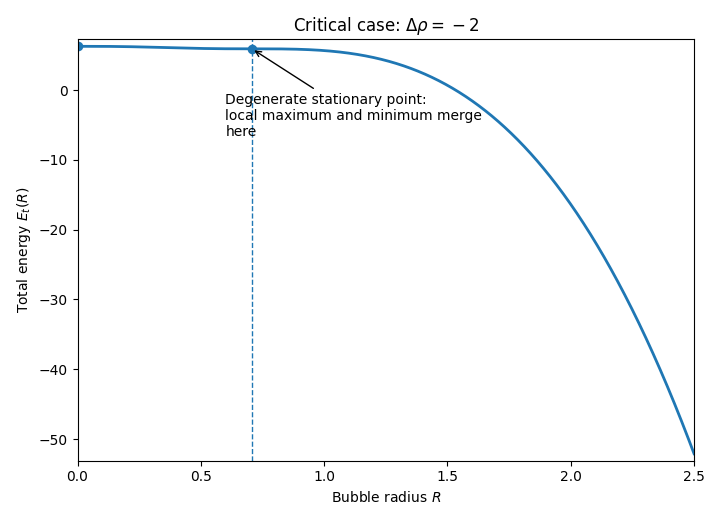}
	\caption{\footnotesize{Illustrative behavior of the total bubble energy $E_{t}(R)$ for the contrast case $\Delta \rho < 0$, generated using the same benchmark values as in Table~\ref{tab:benchmark_energy_plot}. For the choice $\Delta \rho = - 2$, we hit the critical point and therefore the local maximum and minimum merge into a single point at $R \approx 0.7071$, beyond which the negative vacuum term drives runaway expansion.}}
	\label{fig:bubble_energy_negative_case_2}
\end{figure}

\begin{figure}[htbp]
	\centering
	\includegraphics[width=0.65\textwidth]{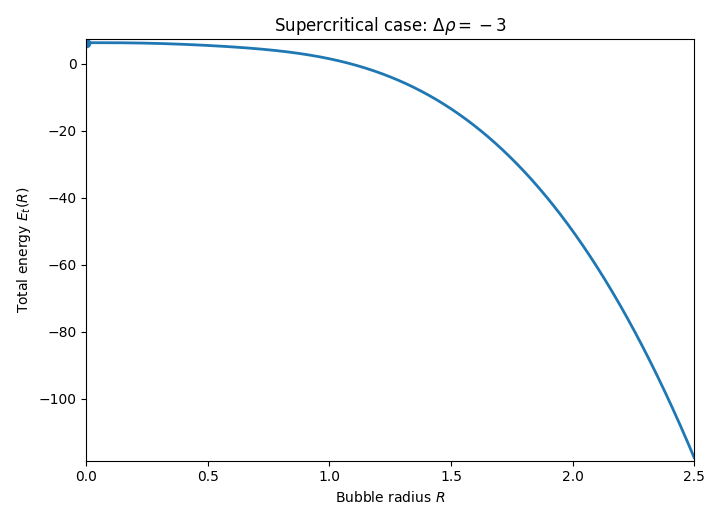}
	\caption{\footnotesize{Illustrative behavior of the total bubble energy $E_{t}(R)$ for the contrast case $\Delta \rho < 0$, generated using the same benchmark values as in Table~\ref{tab:benchmark_energy_plot}. For the illustrative choice $\Delta \rho = -3$, we are in the super-critical regime and therefore no positive stationary point exists. The energy monotonically decreases away from the collapsed configuration at $R = 0$.}}
	\label{fig:bubble_energy_negative_case_3}
\end{figure}

Finally, consider the opposite sign,
\begin{equation}
	\Delta \rho < 0 .
\end{equation}
As emphasized in Sec.~4, this case lies outside the admissible transitions studied in the present work, but it is useful to analyze it as a contrast.
\\ \\
Near the collapsed limit, the small-$R$ expansion of the total energy becomes
\begin{equation}
	E_t(R) = E_t(0) - \frac{4\pi}{3} \, |\Delta \rho| \, R^3 + \frac{2\pi T_{\mathrm{eff}}}{\sqrt{\mathcal{A}}} \, R^4 + O(R^8) .
\end{equation}
The leading correction away from $R = 0$ is therefore negative. Hence the energy decreases as one moves away from the collapsed configuration, and the point $R = 0$ is not a local minimum. In other words, for $\Delta \rho < 0$ the collapsed state is not a stable remnant.
\\ \\
The global structure is determined by
\begin{equation}
	\frac{dE_t}{dR} = 4\pi R^2 \left[ \frac{2 T_{\mathrm{eff}} R}{\sqrt{R^4 + \mathcal{A}}} + \Delta \rho \right] \, .
\end{equation}
Since $\Delta \rho < 0$, it is convenient to define,

\begin{equation}
	b \equiv |\Delta \rho| > 0 ,
\end{equation}
so that,

\begin{equation}
	\frac{dE_t}{dR} = 4\pi R^2 \left[ f(R) - b \right] \, , \qquad f(R) \equiv \frac{2 T_{\mathrm{eff}} R}{\sqrt{R^4 + \mathcal{A}}} \, .
	\label{positivity condition}
\end{equation}
The function $f(R)$ vanishes in both limits,

\begin{equation}
	f(R) \to 0 \quad \text{as} \quad R \to 0 \, , \qquad f(R)\to 0 \quad \text{as} \quad R\to \infty \, ,
\end{equation}
and therefore any positive-radius stationary points can exist only if $f(R)$ becomes larger than $b$ at intermediate radius.
\\ \\
To locate the maximum of $f(R)$, we differentiate,

\begin{equation}
	f'(R) =	\frac{2 T_{\mathrm{eff}}}{(R^4 + \mathcal{A})^{3/2}} \left(\mathcal{A} - R^4\right) = 0 \, .
\end{equation}
Thus $f(R)$ has a single maximum at,

\begin{equation}
	R^4 = \mathcal{A} \, .
\end{equation}
Evaluating $f(R)$ there gives,

\begin{equation}
	f_{\max} = \frac{2 \, T_{\mathrm{eff}} \mathcal{A}^{1/4}}{\sqrt{2 \, \mathcal{A}}} = \frac{\sqrt{2} \, T_{\mathrm{eff}}}{\mathcal{A}^{1/4}} \, .
\end{equation}
Using the values from Table \ref{tab:benchmark_energy_plot}, we obtain the value of $f_{\rm max} (R)$ as,

\begin{equation}
	f_{\max} = \frac{\sqrt{2} \, T_{\mathrm{eff}}}{\mathcal{A}^{1/4}} = 2
\end{equation}
\paragraph{Sub-Critical Case:} From \eqref{positivity condition}, it then follows that positive-radius extrema exist only if,

\begin{equation}
	|\Delta \rho| < f_{\max} \quad \Longrightarrow \quad |\Delta \rho| < 2 \, .
\end{equation}
In that regime, there are two stationary points. Since $\dfrac{dE_t}{dR}<0$ for sufficiently small $R$, it then becomes positive at intermediate radius, and finally becomes negative again at large $R$, the first stationary point is a local minimum and the second is a local maximum. The $\Delta \rho < 0$ branch for $|\Delta \rho| < f_{\max}$ therefore contains a metastable finite-radius bubble separated from runaway expansion by an energy barrier, consult Fig.~\ref{fig:bubble_energy_negative_case}.
\paragraph{Critical Case:} At the critical value,

\begin{equation}
	|\Delta \rho| = 2 \, ,
\end{equation}
the two extrema merge into a single degenerate stationary point at $R^4 = \mathcal{A}$ or $R = \mathcal{A}^{1/4} \approx 0.7071$, see Fig.~\ref{fig:bubble_energy_negative_case_2}.
\paragraph{Super-Critical Case:} If instead,

\begin{equation}
	|\Delta \rho| > 2 \, ,
\end{equation}
then no positive-radius stationary points exist and the energy decreases monotonically for all $R > 0$, see Fig.~\ref{fig:bubble_energy_negative_case_3}.
\\ \\
Finally, note that at large radius, the total energy expression reads,

\begin{equation}
	E_t(R) = 4 \pi T_{\mathrm{eff}} \sqrt{R^4 + \mathcal{A}} + \frac{4\pi}{3}\Delta \rho \, R^3 \,.
\end{equation}
Since $\Delta \rho < 0$, the cubic term is negative and dominates over the DBI wall term at sufficiently large $R$. Therefore the energy eventually decreases without bound, indicating runaway expansion. The $\Delta \rho<0$ branch thus belongs to the vacuum-decay class rather than to the sector of stable collapsed remnants considered in this work.

\FloatBarrier

\section{\centering Interpretation and Classification of the Worldvolume-Flux Supported Remnants}

We are now in a position to interpret the energetically stable collapsed
configurations obtained in Sec.~6 as a new class of localized,
particle-like states supported by quantized worldvolume $U(1)$
monopole flux. The central result of the DBI analysis is that once a membrane electrically charged under the bulk three-form potential $C_3$ also carries a conserved worldvolume $U(1)$ monopole flux, collapse need not end in a trivial zero-energy configuration. Instead, for the physically admissible branch transitions selected by the Hartle--Hawking--Wu feasibility condition \eqref{feasible condition}, the bubble can approach a microscopic collapsed configuration, formally described by the limit \(R\to0\) within the DBI effective theory, while retaining a finite rest energy set by the wall scale and the conserved worldvolume flux. We identify these finite-energy remnants as \emph{topolons}.
\\ \\
This interpretation is physically natural because the remnant is not supported by tension alone. In the absence of monopole flux, the wall energy scales as $E \sim 4 \pi T_{\rm eff} R^{2}$ and therefore vanishes as $R \to 0$, so collapse simply removes the bubble. By contrast, when the worldvolume $U(1)$ carries a nonzero first Chern number, the worldvolume magnetic flux cannot be continuously discharged as the wall shrinks. The full DBI dynamics then regulate the short-distance behavior and replace the would-be trivial endpoint by a finite-mass collapsed state. The stability of the remnant therefore arises from the combined effect of topology, flux conservation, and nonlinear brane dynamics.
\\ \\
It is useful to classify these objects according to the sign of the vacuum-energy difference $\Delta \rho$ between the interior and exterior branches. At the level of the effective energy functional, three qualitatively distinct classes may be identified. However, only the first two classes lie within the HHW-admissible four-form flux interval defined in Sec.~4. The third is retained only for comparison, because it is not prepared by the compact \(S^4\) no-boundary saddle adopted in the present semiclassical construction.
\\ \\
We define \textbf{Class I topolons} to be the flux supported remnants associated with,

\begin{equation}
	\Delta \rho = 0 \, .
\end{equation}
These correspond to the exact sign-flip channel, for which the daughter and parent branches have degenerate vacuum energies. Class I topolons are therefore the cleanest realization of a stable collapsed remnant, since their mass is generated entirely by the microscopic wall sector and the conserved worldvolume magnetic flux, with no residual vacuum-pressure contribution.
\\ \\
We define \textbf{Class II topolons} to be the worldvolume magnetic flux supported remnants associated with
\begin{equation}
	\Delta \rho > 0 \, .
\end{equation}
These correspond to the more generic transitions between branches within the Hartle--Hawking--Wu admissible four-form flux interval. In this case the interior vacuum energy is higher than the exterior one, so the total bubble energy contains a positive volume contribution proportional to $R^{3}\Delta \rho$. Even so, the small-radius behavior remains qualitatively the same as in Class I. The energy is still minimized in the collapsed limit, and the collapsed configuration remains an energetically stable, finite-mass, particle-like state within the static spherical DBI
description.
\\ \\
For contrast, one may also define \textbf{Class III bubbles}, corresponding to,

\begin{equation}
	\Delta \rho < 0 \, .
\end{equation}
These do not belong to the stable topolon sector. As shown in Sec.~6, their
energy profile differs qualitatively from that of Class~I and Class~II, since the collapsed configuration is not a local minimum and the large-radius behavior is driven toward runaway expansion. The Class~III sector further splits into three regimes. For $|\Delta \rho| < \sqrt{2} \, T_{\rm eff}/\mathcal{A}^{1/4}$, the energy contains a local minimum and a local maximum, so the bubble is metastable and separated from runaway expansion by an energy barrier. This is the regime for which a Euclidean
instanton or bounce analysis, followed by Lorentzian bubble growth, would be
the natural framework for determining the decay rate. At the critical value $|\Delta \rho| = \sqrt{2} \, T_{\rm eff}/\mathcal{A}^{1/4}$, these two extrema merge into a single degenerate stationary point at $R^4 = \mathcal{A}$, marking the threshold between metastability and runaway growth. For $|\Delta \rho| > \sqrt{2} \, T_{\rm eff}/\mathcal{A}^{1/4}$, no positive-radius stationary point exists and the energy decreases monotonically for all $R > 0$, so the bubble directly enters the runaway vacuum-decay channel. Class III bubbles should therefore be regarded as decay channels, not as particle states.
\\ \\
This distinction is physically important. Class I and Class II configurations behave, after formation, as ordinary heavy localized states. Their defining property is that the total energy is minimized in the collapsed limit, so once produced they need not expand nor relax to a trivial zero-energy configuration. Instead, they persist as finite-mass remnants. This is precisely why a Euclidean instanton analysis is not required for these two classes of topolons. Moreover, our purpose is not to compute a branch-changing decay rate, but to establish the existence and stability of the flux supported remnants themselves. Once such a remnant exists and is energetically stable, it is natural to treat it, for cosmological purposes, as a particle species. Its cosmological production may then be modeled, at the level of coarse-grained relic abundances, using methods similar to those employed for a heavy nonrelativistic particle species. This analogy however does not imply that the topolon is microscopically an elementary scalar field. A detailed analysis of topolon production in the early universe and their resulting cosmological abundance is deferred to future work.
\\ \\
A further useful feature of the present construction is that the HHW
admissibility condition excludes the corresponding
$\Delta\rho < 0$ daughter branches from the semiclassical branch set
considered here. It therefore restricts the retained transitions to the
$\Delta\rho \geq 0$ classes for which the static spherical DBI energy
functional is minimized at the collapsed endpoint. Once the DBI
membrane theory and its conserved worldvolume monopole flux are
specified, no further ad hoc stabilizing potential is required.
\\ \\
The resulting finite-energy remnants define a new class of microscopic,
particle-like objects, which we call \emph{topolons}. Their general
collapsed-state masses are,

\begin{equation}
	E_n \; = \; \frac{2 \, \pi \, |n| \, \sqrt{T_{\rm eff}}}{g_{\rm YM}} \,,
\end{equation}
and reduce to,

\begin{equation}
	E_n \; = \; 2 \, \pi \, |n| \, T_{\rm eff}^{1/3}
\end{equation}
under the illustrative benchmark relation \eqref{eq:gym_benchmark}. Their energetic stability follows from the shape of the full DBI energy functional together with the restriction to HHW-admissible four-form flux transitions. In this sense, topolons are naturally interpreted as topological, brane-like relics of the three-form--membrane system.

\section{\centering Discussion and Limitations}

We begin with the first important caveat, that is the DBI action is itself an effective field theory. It captures the low-energy worldvolume dynamics of a brane-like membrane, including the tension term, the abelian worldvolume gauge field, and the leading nonlinear completion of the gauge sector. It is therefore a natural and economical framework for studying whether conserved flux can obstruct complete collapse. However, it should not be regarded as a complete ultraviolet description of the membrane microphysics. In a more complete theory, additional operators may appear, including higher-derivative corrections and corrections associated with heavy modes that have been integrated out. Such effects may shift the precise quantitative relation between the remnant mass, the wall tension, the gauge coupling, and the flux number. What the present analysis shows is that within the minimal DBI effective theory, the combination of conserved monopole flux and nonlinear worldvolume dynamics is already sufficient to produce a finite-energy collapsed endpoint.
\\ \\
This point is closely related to the interpretation of the limit $R \to 0$. In the effective theory, this limit should not be read literally as the formation of an object of exactly vanishing size. Rather, it means that the energetically preferred radius is driven below all macroscopic scales resolved by the low-energy description. Any ultraviolet completion is expected to introduce a microscopic length scale that regulates the short-distance core and prevents the object from becoming truly pointlike. The same conclusion is also motivated by gravity. A strictly pointlike object of fixed nonzero mass would produce arbitrarily large curvature in its immediate vicinity, which signals the breakdown of the low-energy description. It is therefore more physical to assume that the ultraviolet completion, or possibly self-gravity itself, resolves the object into a finite core. To make this explicit, we introduce an effective core scale
\begin{equation}
	\ell_{\rm core} \;\equiv\; \max \bigl(\ell_{\rm Pl}, \ell_{\rm UV}\bigr) \, , \qquad \ell_{\rm Pl} = \sqrt{G} = \frac{1}{M_{\rm Pl}}
\end{equation}
so that,

\begin{equation}
	r_{\rm core} \sim \ell_{\rm core} \, .
\end{equation} 
On probe scales $L_{\rm probe} \gg r_{\rm core}$, the object is effectively pointlike and finite-size effects can be encoded by higher-dimension operators suppressed by powers of $r_{\rm core}/L_{\rm probe}$. In this sense, the collapsed topolon should be understood exactly as one understands any other effective point particle, namely as an object whose internal structure is unresolved on the scales relevant to the dynamics of interest. The gravitational collapse of the topolon into a black hole is avoided provided its Schwarzschild radius remains parametrically smaller than the core size,

\begin{equation}
	r_{s}(E_{n}) \ll \ell_{\rm core} \, ,
\end{equation}
with,

\begin{equation}
	r_{s}(E_{n}) = 2 G E_{n} \, ,
\end{equation}
where $E_n$ is given by equation \eqref{En definition}. A detailed analysis of the effective core size of topolons and their gravitational back-reaction will be presented alongside the study of their cosmological production.
\\ \\
A second limitation concerns the status of the admissible flux interval. In the present framework, the restriction to the sector $q \in [-q_{0}, +q_{0}]$ is not derived as a theorem of local bulk dynamics alone. Rather, it is motivated by the Hartle--Hawking--Wu semiclassical selection argument and interpreted as a restriction on the set of semiclassically realizable branches. This is sufficient for the internal logic of the present construction at a semi-classical level, but it also means that the admissible sector should be viewed as a cosmological input to the effective theory rather than as a purely local dynamical result. A more complete treatment would require a deeper derivation of the branch-selection principle, possibly within a fuller quantum cosmological or ultraviolet framework.
\\ \\
A related idealization is that we have worked with a single three-form sector and have not attempted to explain the observed small late-time vacuum energy of the Universe from this sector alone. The ambient branch is taken to satisfy $\Lambda_{\rm eff}(q_{\rm out}) \simeq 0$ as a controlled idealization, with the understanding that any residual vacuum energy may arise from additional sectors or mechanisms not modeled here. Likewise, the benchmark choices for the \textit{membrane charge spacing} $q_{\rm min}$ and the \textit{worldvolume coupling} $g_{\rm YM}$ are representative possibilities rather than universal predictions of the framework. The present paper is therefore best understood as isolating one dynamical question, namely whether a flux-carrying membrane bubble can collapse to a stable finite-energy remnant, rather than as providing a complete microscopic model of the cosmological constant or of ultraviolet membrane physics.
\\ \\
Furthermore, configurations with $\Delta \rho < 0$ correspond to runaway expanding bubbles and belong to the usual first-order vacuum-decay or the direct vacuum-decay channel. The present paper does not compute Euclidean bounce actions, nucleation rates, or Lorentzian post-nucleation evolution for that class. Those observables are relevant for Class III bubbles, but they address a different physical problem from the one studied here. In this sense, the present work is a stability analysis of the collapsed endpoint, not a full semiclassical treatment of all possible membrane nucleation channels. We may study Class III bubbles in detail in a future work.
\\ \\
The cosmological implications are also worth mentioning here. Since Class I and Class II topolons behave as stable heavy particle species after formation, the relevant abundance question is not the usual false-vacuum decay problem but rather the production of heavy nonperturbative relics in the early Universe. Depending on the thermal history and on the time dependence of the background curvature, such objects may be produced gravitationally, thermally, or nonthermally. Determining the dominant channel requires a dedicated analysis of the production rate, the resulting momentum distribution, the late-time relic abundance, and possible self-interactions. These quantitative issues are left for future work as mentioned earlier. The present paper identifies the stable microscopic sector that such a cosmological study must build upon.
\\ \\
Furthermore, at long distances, topolons may behave as heavy particle-like states if they possess no significant long-range interactions other than gravity. This makes them natural dark-matter candidates. In particular, sufficiently massive topolons produced in the early universe would behave as cold nonrelativistic relics at late times. Whether this possibility is realized in detail depends on the production history, the resulting mass spectrum, and the abundance of the stable remnants. A full cosmological analysis lies beyond the scope of the present work, but the classification developed here already identifies the stable sector that should be carried forward into such a study.
\\ \\
With these caveats in mind, the main message remains robust. Within the minimal DBI effective theory, conserved worldvolume monopole flux changes the endpoint of collapse in an essential way. Instead of disappearing into a trivial zero-energy state, Class I and Class II flux-carrying bubbles (topolons) can relax to microscopic but finite-energy remnants, whereas Class III bubbles are unstable and undergo runaway expansion. The topolon should therefore be viewed as an emergent brane-like relic of the underlying flux sector, whose precise core structure is ultraviolet-sensitive but whose long-distance particle-like behavior is already visible at the level of the effective theory.
\\ \\
Finally, the broader significance of the present construction lies in the
economy of the underlying three-form gauge structure. At the
homogeneous level, the constant flux of the four-form field strength
labels vacuum branches and may participate in neutralizing the dominant
permanent vacuum-energy contribution. This operation concerns the
non-inflationary vacuum branch and does not require the total energy of
the Universe to vanish. A dynamical scalar field may subsequently
supply a temporary positive contribution along a no-boundary saddle and
thereby generate inflationary initial conditions without undoing the
neutralization of the permanent branch. A small residual late-time
vacuum energy may then arise through further four-form dynamics,
additional sectors, infrared gravitational effects, or presently
unknown physics.
\\ \\
The present work identifies a complementary, nonhomogeneous
manifestation of the same gauge structure. Membranes electrically
charged under the three-form potential interpolate between distinct
four-form flux branches. When such a membrane also carries an
independent quantized worldvolume \(U(1)\) monopole flux, its nonlinear
DBI dynamics can convert the collapsed endpoint from a trivial
zero-energy configuration into a finite-energy particle-like remnant.
Thus, the homogeneous four-form flux controls the vacuum branch, while
nonperturbative charged-membrane configurations of the same three-form
gauge sector can supply a candidate dark relic.
\\ \\
These observations suggest a possible unified program in which
vacuum-energy neutralization, inflationary initial conditions, residual
dark energy, and dark matter are organized around a common three-form
gauge sector. The word ``unified'' should, however, be understood in a
structural rather than reductive sense. The inflaton remains an
additional dynamical scalar; the membrane tension and its worldvolume
\(U(1)\) field belong to an independent brane effective theory; and
the origin of the observed residual vacuum energy is not fixed by the
present analysis.

\section{\centering Conclusion}

In this work we have studied membrane-mediated branch transitions in a three-form flux sector and shown that the endpoint of bubble collapse can be qualitatively altered once the wall carries a conserved worldvolume monopole flux and is described by the full Dirac--Born--Infeld action. In the absence of such flux, a collapsing spherical wall relaxes to a trivial zero-energy configuration. By contrast, when the wall supports a nonzero first Chern number, the nonlinear DBI dynamics regulate the short-distance behavior and yield a finite nonzero rest energy in the collapsed limit. This provides a simple mechanism through which a flux-carrying vacuum bubble can terminate not in disappearance, but in a stable microscopic remnant.
\\ \\
We have interpreted these finite-energy collapsed objects as a new class of particle-like states, which we call topolons. Their existence does not rely on the introduction of an additional ad hoc stabilizing sector. Rather, it follows directly from the interplay between quantized worldvolume flux, membrane tension, and the nonlinear structure of the DBI action. In this sense, the topolon is not an elementary degree of freedom inserted by hand, but an emergent brane-like relic of the underlying flux sector.
\\ \\
A central role is played by the Hartle--Hawking--Wu feasibility condition, which restricts the semiclassically admissible branch transitions to the interval $q \in [-q_{0}, +q_{0}]$. Within this admissible sector, the effective energy analysis leads naturally to two stable classes of flux-supported remnants and one contrasting unstable class. Class I topolons correspond to $\Delta \rho = 0$ and represent the cleanest realization of a stable collapsed remnant, with mass arising purely from the microscopic wall sector and the conserved flux. Class II topolons correspond to $\Delta \rho > 0$ and remain stable finite-mass remnants even in the presence of a positive vacuum-energy contribution to the bubble energy. By contrast, configurations with $\Delta \rho < 0$ belong to a different dynamical channel, namely runaway vacuum decay, and should be treated as expanding decay bubbles rather than as particle states. The HHW restriction is therefore important not only for branch selection, but also because it isolates the stable remnant sector from the unwanted decay channel.
\\ \\
The present analysis should nevertheless be viewed as an effective one. The DBI action provides a controlled low-energy description of the worldvolume dynamics, while the collapsed limit should be interpreted as a statement that the preferred radius is driven below all macroscopic probe scales, not that the object becomes literally of zero size. A more complete ultraviolet treatment is expected to resolve the core at a finite microscopic scale and may modify the detailed near-core structure. Likewise, sufficiently large masses or flux numbers may require a treatment including gravitational backreaction. These issues do not alter the central qualitative result established here, namely that conserved worldvolume flux can convert the endpoint of collapse from a trivial disappearance into a finite-energy stable remnant.
\\ \\
The result also acquires a broader interpretation when viewed together
with the homogeneous role of the three-form gauge sector. The constant
four-form flux may select a branch on which the dominant permanent
vacuum-energy contribution is neutralized, while a dynamical scalar may
in principle provide the finite temporary energy required for an
inflationary no-boundary history. The present construction shows that
the same underlying gauge sector also possesses a nonperturbative
charged-membrane component whose collapsed configurations can survive as
finite-energy particle-like remnants. A small residual late-time vacuum
energy may arise from additional four-form dynamics, another sector, or
physics not included in the present effective description.
\\ \\
The emerging picture is therefore not that a bare three-form field
alone explains every cosmological component, but that a single
three-form gauge sector can provide an economical common backbone for
vacuum-energy neutralization, inflationary initial conditions, residual dark energy and nonperturbative dark-relic formation. The present paper establishes the microscopic topolon and stability component of this program. Whether this structure can be completed into a quantitatively viable unified framework for the dark sector depends on the inflationary saddle, the origin of the residual vacuum energy, the cosmological production and abundance of topolons, and the ultraviolet and gravitational resolution of their collapsed cores.

\section*{\centering Declarations}

\subsection*{Ethics Declaration}
Ethics declaration: not applicable. This study is theoretical in nature and does not involve human participants, human data, animals, or biological materials.

\subsection*{Consent to Participate}
Consent to Participate declaration: not applicable.

\subsection*{Consent to Publish}
Consent to Publish declaration: not applicable.

\subsection*{Data Availability}
Data Availability declaration: no datasets were generated or analyzed during the current study. All calculations are analytical and all relevant derivations are included in the manuscript.

\subsection*{Author Contribution}
The author confirms sole responsibility for the conception and design of the work, the analytical calculations, the interpretation of the results, the preparation of the manuscript, and the final approval of the submitted version.

\subsection*{Competing Interests}
The author declares no competing interests.

\newpage

\appendix

\section{\centering The Stress-Energy Tensor of $C_3$}

The general formula for the matter stress–energy tensor is given by,

\begin{equation}
	T_{\mu\nu} = -\frac{2}{\sqrt{-g}} \; \frac{\delta S_{\text{matter}}}{\delta g^{\mu\nu}} 
	\label{Stress Energy Tensor}
\end{equation}
Varying the action $S_3$ with respect to the metric gives,

\begin{equation}
	\delta S_{3} = -\frac{1}{2\!\cdot\!4!}\int d^{4}x
	\left[\, (\delta\sqrt{-g}) \, F_{\alpha\beta\gamma\delta} F^{\alpha\beta\gamma\delta} + \sqrt{-g} \, F_{\alpha\beta\gamma\delta} \, \delta F^{\alpha\beta\gamma\delta}
	\right]
	\label{S3 Variation}
\end{equation}
Using the identity,

\begin{equation}
	\delta\sqrt{-g}= -\frac12\sqrt{-g}\;g_{\mu\nu}\,\delta g^{\mu\nu}
\end{equation}
The variation of the action $S_3$ becomes,

\begin{equation}
	\delta S_3 = -\frac{1}{2\!\cdot\!4!}\int d^{4}x \; \sqrt{-g} \;
	\Bigl[-\frac12 g_{\mu\nu} \; \delta g^{\mu\nu} F_{\alpha\beta\gamma\delta} F^{\alpha\beta\gamma\delta} + \,F_{\alpha\beta\gamma\delta} \,\delta F^{\alpha\beta\gamma\delta}
	\Bigr]
\end{equation}
Expressing $F^{\alpha\beta\gamma\delta} = g^{\alpha\kappa} g^{\beta\lambda} g^{\gamma\mu} g^{\delta\nu} F_{\kappa\lambda\mu\nu}$ and varying one metric inverse at a time, we get,

\begin{align}
	\delta F^{\alpha\beta\gamma\delta}
	& = \phantom{+\,} (\delta g^{\alpha\kappa}) g^{\beta\lambda} g^{\gamma\mu} g^{\delta\nu} F_{\kappa\lambda\mu\nu}
	+ g^{\alpha\kappa} (\delta g^{\beta\lambda}) g^{\gamma\mu} g^{\delta\nu}
	F_{\kappa\lambda\mu\nu} \\ \nonumber
	& \quad + g^{\alpha\kappa} g^{\beta\lambda} (\delta g^{\gamma\mu}) g^{\delta\nu} F_{\kappa\lambda\mu\nu} + g^{\alpha\kappa} g^{\beta\lambda} g^{\gamma\mu} (\delta g^{\delta\nu}) F_{\kappa\lambda\mu\nu}
	\label{Master Equation}
\end{align}
Consider the first term $(\delta F)_1$ of $\delta F^{\alpha\beta\gamma\delta}$,

\begin{equation}
	(\delta F)_1 = (\delta g^{\alpha\kappa}) g^{\beta\lambda} g^{\gamma\mu} g^{\delta\nu} F_{\kappa\lambda\mu\nu}
\end{equation}
Performing the metric contractions with $F_{\kappa\lambda\mu\nu}$, we get,

\begin{equation}
	(\delta F)_1 = \delta g^{\alpha\kappa} F_{\kappa}^{\; \ \beta\gamma\delta}
\end{equation}
The algebraic structure of the other three terms in equation (A.5) is exactly the same and we therefore get,

\begin{equation}
	\delta F^{\alpha\beta\gamma\delta} = 4 \, F_{\mu}^{\; \ \beta\gamma\delta} \delta g^{\mu \alpha}
\end{equation}
Therefore equation \eqref{S3 Variation} becomes,

\begin{equation}
	\delta S_3 = -\frac{1}{2 \cdot 4!} \int d^{4}x \; \sqrt{-g} \;\Bigl[
	-\frac{1}{2} g_{\mu\nu} F_{\alpha\beta\gamma\delta} F^{\alpha\beta\gamma\delta} + 4 \, F_{\nu\beta\gamma\delta} \, F_{\mu}^{\; \ \beta\gamma\delta} \Bigr] \delta g^{\mu\nu}
\end{equation}
Comparing the equation above with \eqref{Stress Energy Tensor}, we see that the stress-energy tensor $T_{\mu\nu}^{(q)}$ of the four-form sector is,

\begin{equation}
	T_{\mu\nu}^{(q)} = \frac{1}{4!} \Bigl[ 4 \, \,F_{\nu\beta\gamma\delta} \, F_{\mu}^{\; \ \beta\gamma\delta} - \frac{1}{2} g_{\mu\nu} F_{\alpha\beta\gamma\delta} F^{\alpha\beta\gamma\delta}
	\Bigr]
\end{equation}
Using $F_{\alpha\beta\gamma\delta} = q \, \epsilon_{\alpha\beta\gamma\delta}$ in the equation above, we get,

\begin{equation}
	T_{\mu\nu}^{(q)} = - q^2 g_{\mu\nu} + \frac12 q^2 g_{\mu\nu}
\end{equation}
That is,

\begin{equation}
	T_{\mu\nu}^{(q)} = -\frac12 q^2 g_{\mu\nu}
\end{equation}
We therefore see that the stress-energy tensor of the four-form sector is exactly of the form of a fluid with equation of state $P_{q} = -\rho_{q}$.

\newpage

\section{\centering The Four-Form Flux Jump Condition}

We begin from the action for $C_3$ in the presence of sources, denoted by $S_3$, which reads,

\begin{equation}
	S_3 \, = \, - \, \frac{1}{2 \cdot 4!} \, \int d^4 x \, \sqrt{-g} \, F_{\mu\nu\rho\sigma} F^{\mu\nu\rho\sigma} \; + \; q_b \, \int_{\Sigma_3} C_3
	\label{S3 coupled 0}
\end{equation}
where $q_b$ is the charge of the bubble. Let $x^{\mu} \, = \, (t,r,\theta,\phi)$ be the coordinates of the target or embedding spacetime and $\xi^{\alpha} \, = \, (t,\theta,\phi)$ be the coordinates on the bubble's worldvolume. We assume the bubbles have a radius $R$ and therefore, the embedding of these bubbles in spacetime is simply given by $X^{\mu}(\xi^{\alpha}) \, = \, (t,R,\theta,\phi)$. With this embedding, the pullback of $C_3$ on the bubble's worldvolume reads,

\begin{equation}
	\left. C_3 \right|_{\Sigma_3} = \left( \frac{1}{3!} \, \epsilon^{abc} \,
	\frac{\partial X^{\mu}}{\partial \xi^{a}} \, \frac{\partial X^{\nu}}{\partial \xi^{b}} \, \frac{\partial X^{\rho}}{\partial \xi^{c}} \, C_{\mu\nu\rho}(X(\xi)) \right) \, d\xi^{0} \wedge d\xi^{1} \wedge d\xi^{2}
\end{equation}
and the action $S_3$ becomes,

\begin{equation}
	S_3 = - \, \frac{1}{2 \cdot 4!} \, \int d^4 x \, \sqrt{-g} \, F_{\mu\nu\rho\sigma} F^{\mu\nu\rho\sigma} + \frac{q_b}{3!} \, \int_{\Sigma_3} d^3 \xi \, \epsilon^{abc} \frac{\partial X^{\mu}}{\partial \xi^{a}} \, \frac{\partial X^{\nu}}{\partial \xi^{b}} \, \frac{\partial X^{\rho}}{\partial \xi^{c}} \, C_{\mu\nu\rho}(X(\xi))
	\label{S3 action coupled}
\end{equation}
Notice that we can also express the interaction term $S_{\rm int}$ using the contravariant 3-current density tensor $J^{\mu\nu\rho}$ as,

\begin{equation}
	S_{\rm int} \, = \, \frac{1}{3!} \, \int d^4 x \, \sqrt{-g(x)} \, J^{\mu\nu\rho}(x) \, C_{\mu\nu\rho}(x)
	\label{S3 action coupled 2}
\end{equation}
Comparing equations (B.3) and \eqref{S3 action coupled 2}, we find,

\begin{equation}
	J^{\mu\nu\rho} = q_b \, \int d^3 \xi \, \frac{\epsilon^{abc}}{\sqrt{-g(x)}} \, \frac{\partial X^{\mu}}{\partial \xi^{a}} \, \frac{\partial X^{\nu}}{\partial \xi^{b}} \, \frac{\partial X^{\rho}}{\partial \xi^{c}} \, \delta^{(4)}(x - X)
\end{equation}
where $\delta^{(4)}(x - X)$ is the four-dimensional Dirac delta, given by,

\begin{equation}
	\delta^{(4)}(x - X) = \delta(t - \xi^{0}) \, \delta(r - R) \, \delta(\theta - \xi^{1}) \, \delta(\phi - \xi^{2})
\end{equation}
Varying the interaction term as defined by equation \eqref{S3 coupled 0}, we get,

\begin{equation}
	\delta S_{\rm int} = \frac{1}{3!} \, \int d^4 x \, \sqrt{-g(x)} \, J^{\mu\nu\rho}(x) \, \delta C_{\mu\nu\rho}(x)
\end{equation}
and setting $\delta S \, = \, \delta S_{\rm gauge} + \delta S_{\rm int} \, = \, 0$, we get the equation of motion of $F_4$ as,

\begin{equation}
	\nabla_{\mu} F^{\mu\nu\rho\sigma} \, = \, J^{\nu\rho\sigma}
\end{equation}
That is,

\begin{align}
	\frac{1}{\sqrt{-g(x)}} & \, \partial_{\mu} \big( \sqrt{-g(x)} \, F^{\mu\nu\rho\sigma} \big) \\ \nonumber
	& = q_b \, \int d^3 \xi \, \frac{1}{\sqrt{-g(x)}} \, \epsilon^{abc} \frac{\partial X^{\nu}}{\partial \xi^{a}} \, \frac{\partial X^{\rho}}{\partial \xi^{b}} \, \frac{\partial X^{\sigma}}{\partial \xi^{c}} \, \delta^{(4)}(x - X)
	\label{Jump 1}
\end{align}
Note that,

\begin{equation}
	\epsilon^{abc} \frac{\partial X^{\nu}}{\partial \xi^{a}} \,
	\frac{\partial X^{\rho}}{\partial \xi^{b}} \, \frac{\partial X^{\sigma}}{\partial \xi^{c}} = \epsilon^{t\theta\phi} \, \delta^{\nu}_{t} \, \delta^{\rho}_{\theta} \, \delta^{\sigma}_{\phi}
\end{equation}
and therefore equation (B.9) reduces to,

\begin{equation}
	\partial_{\mu} \big( \sqrt{-g(x)} \, F^{\mu t \theta \phi} \big) =
	q_b \, \int d^3 \xi \, \epsilon^{t\theta\phi} \, \delta^{(4)}(x - X)
\end{equation}
The integral $\int_{\Sigma_3} d^3 \xi \, \epsilon^{t\theta\phi} \, \delta(t - \xi^{0}) \, \delta(\theta - \xi^{1}) \, \delta(\phi - \xi^{2}) \, = \, 1$ and we therefore have,

\begin{equation}
	\partial_{\mu} \big( \sqrt{-g(x)} \, F^{\mu t \theta \phi} \big) \, = \, q_b \, \delta(r - R)
\end{equation}
The general solution of $F_4$ takes the form $F^{\mu t \theta \phi} \, = \, q \, \epsilon^{\mu t \theta \phi}$. Therefore, the equation above reads,

\begin{equation}
	\partial_{\mu} \big( \sqrt{-g(x)} \, q \, \epsilon^{\mu t \theta \phi} \big) \, = \, q_b \, \delta(r - R)
\end{equation}
The Levi-Civita tensor $\epsilon^{\mu t \theta \phi}$ constrains the choice for $\mu$ to be $r$ (as it is zero for any other combination of indices) and we therefore get,

\begin{equation}
	\partial_{r} \big( \sqrt{-g(x)} \, q \, \epsilon^{r t \theta \phi} \big) \, = \, q_b \, \delta(r - R)
\end{equation}
As $\epsilon^{r t \theta \phi} \, = \, + \, \dfrac{1}{\sqrt{-g}}$, we have,

\begin{equation}
	\partial_{r} q \, = \, q_b \, \delta(r - R)
\end{equation}
Integrating both sides gives,

\begin{equation}
	\int_{q_{\rm in}}^{q_{\rm out}} dq = q_b \, \lim_{\epsilon \rightarrow 0} \, \int_{R - \epsilon}^{R + \epsilon} \delta(r - R) \, dr
\end{equation}
This gives the jump condition,

\begin{equation}
	q_{\rm out} - q_{\rm in} \, = \, q_b
\end{equation}

\newpage

\section{\centering Effective Cosmological Constant}

Consider the following action describing the dynamics of gravity and a three-form potential $C_3$, with $C_3$ fixed on the equator $\Sigma_{\ast}$ (that is $\delta C|_{\Sigma_{\ast}} \, = \, 0$) which Hawking examined \cite{Hawking1984},

\begin{equation}
	S_E[g,C] = - \, \int d^4 x \, \sqrt{g} \left[ \frac{1}{16\pi G} (R - 2\Lambda_0)	- \frac{1}{48} \, F_{\mu\nu\rho\sigma} F^{\mu\nu\rho\sigma} \right] \, , \qquad F \, = \, dC \, .
	\label{HH with FF}
\end{equation}
This boundary choice is the root of the problem but let us proceed anyway.

\subsection{\centering Hawking's Approach}

Note that equation \eqref{HH with FF} can equivalently be expressed in the language of differential forms as,

\begin{equation}
	S_E[g,C] = - \, \int \frac{1}{16\pi G} (R - 2\Lambda_0) \, {\rm vol}_4
	+ \frac{1}{2} \int_{M} F \wedge \star F \, .
	\label{HH4}
\end{equation}
The no-boundary wavefunction $\Psi_0$ with the above action then reads,

\begin{equation}
	\Psi_q[h_{ij}] = \int_{g|_{\Sigma_{\ast}} \, = \, h \, , \, C|_{\Sigma_{\ast}} \, = \, C_{\Sigma_{\ast}}} \mathcal{D}g \, \mathcal{D}C \, \exp(-S_E[g,C]) \, .
\end{equation}
The leading contribution to the path integral occurs at the stationary configuration $\bar{g}, \bar{C}$ that satisfy $\delta S_E[\bar{g},\bar{C}] \, = \, 0$ with the stated boundary condition. The path integral at leading order is then $\Psi_q[h_{ij}] \, \simeq \, \exp \big[ -S_E(\bar{g},\bar{C}) \big]$ times a one-loop determinant. Hawking then performed the saddle by first integrating out the four form by varying $S_E[g,C]$ with respect to $C$ at fixed $g$ with boundary condition $\delta C|_{\Sigma_{\ast}} \, = \, 0$, which resulted in,

\begin{equation}
	\left. \delta S_E[g,C] \right|_{C} = \int_{M} \delta C \wedge d \star F \, .
\end{equation}
Setting $\left. \delta S_E[g,C] \right|_{C} \, = \, 0$, the field equations for $F_4$ are then easy to read,

\begin{equation}
	d \star F \, = \, 0	\quad \Longleftrightarrow \quad	\nabla_{\mu} F^{\mu\nu\rho\sigma} \, = \, 0 \, ,
\end{equation}
with the solution $F^{\mu\nu\rho\sigma} \, = \, q \, \dfrac{\tilde{\epsilon}^{\mu\nu\rho\sigma}}{\sqrt{g}}$ (see Sec.~2). Hawking then proceeded to substitute the result of the $C_3$ saddle back into $S_E[g,C]$ and obtained,

\begin{equation}
	S_E[g,\bar{C}] = - \, \int d^4 x \, \sqrt{g} \left[ \frac{1}{16\pi G} (R - 2\Lambda_0) - \frac{1}{48} \, q^2 \, \epsilon_{\mu\nu\rho\sigma} \, \epsilon^{\mu\nu\rho\sigma} \right] \, .
\end{equation}
In Euclidean signature $(+,+,+,+)$, the contraction of the Levi-Civita tensor is $\epsilon_{\mu\nu\rho\sigma} \, \epsilon^{\mu\nu\rho\sigma} \, = \, 4!$ and therefore,

\begin{equation}
	S_E[g,\bar{C}] = - \, \frac{1}{16\pi G} \, \int d^4 x \, \sqrt{g} \left[
	R - 2(\Lambda_0 + 4\pi G q^2) \right] \, .
\end{equation}
Defining $\Lambda_{\rm eff} \, = \, \Lambda_0 + 4\pi G q^2$, Hawking then performed the metric saddle and found that the dominant contribution to $\Psi_q$ comes from the following classical solution (subject to the boundary data),

\begin{equation}
	G_{\mu\nu} + \Lambda_{\rm eff} g_{\mu\nu} = G_{\mu\nu} + \left(\Lambda_0 + 4\pi G q^2\right)g_{\mu\nu} = 0 \, .
	\label{Hawking's Effective Lambda}
\end{equation}
Therefore, Hawking arrives at an effective cosmological constant which reads $\Lambda_{\rm eff} = \Lambda_0 + 4\pi G q^2$ as claimed in the main text.

\subsection{\centering Duff’s Criticism}

However, Duff \cite{Duff1989} pointed out that Hawking’s ``substitute the on-shell value of $F_4$ and then vary'' step was incorrect for constrained systems. Rather than substituting first, if one had instead varied the four-form piece and then substituted its on-shell value, the effective cosmological constant would read $\Lambda_{\rm eff} \, = \, \Lambda_0 - 4\pi G q^2$ in contrast to Hawking’s $\Lambda_0 + 4\pi G q^2$. This is immediately obvious on taking the variation of $\displaystyle \frac{1}{2} \int_{M} F \wedge \star F$ with respect to the metric $g_{\mu\nu}$, which gives,

\begin{equation}
	T^{(q)}_{\mu\nu} = \frac{1}{6} \, F_{\mu\alpha\beta\gamma} \, F_{\nu}{}^{\alpha\beta\gamma} - \frac{1}{48} \, g_{\mu\nu} \, F_{\alpha\beta\gamma\delta} \, F^{\alpha\beta\gamma\delta}
\end{equation}
and after substituting the on-shell value of $F_4$ (in Euclidean metric signature) results in,

\begin{equation}
	T^{(q)}_{\mu\nu} \, = \, \frac{1}{2} \, q^2 \, g_{\mu\nu} \, .
\end{equation}
As a result, the Euclidean gravitational field equations evaluated on the metric saddle then read,

\begin{equation}
	G_{\mu\nu} + \Lambda_0 g_{\mu\nu} = 8\pi G \, T^{(q)}_{\mu\nu} = 4\pi G q^2 \, g_{\mu\nu} \; \; \Longrightarrow \; \; G_{\mu\nu} + (\Lambda_0 - 4\pi G q^2) g_{\mu\nu} \, = \, 0 \, .
	\label{Effective Lambda of Duff}
\end{equation}
Therefore, Duff showed that the effective cosmological constant should read $\Lambda_{\rm eff} = \Lambda_0 - 4\pi G q^2$ as opposed to Hawking's $\Lambda_{\rm eff} = \Lambda_0 + 4\pi G q^2$.

\subsection{\centering Wu's Resolution}

Wu switched the variational problem from fixed potential $\delta C_3|_{\Sigma_{\ast}} \, = \, 0$ to fixed flux $(\delta(\star F))|_{\Sigma_{\ast}} \, = \, 0$ \cite{Wu2008}. This required the addition of a Legendre boundary term $\displaystyle - \int_{\Sigma_{\ast}} C \wedge \star F$ in equation \eqref{HH4} as shown,

\begin{equation}
	S_E[g,C] = - \, \int \frac{1}{16\pi G} (R - 2\Lambda_0) \, {\rm vol}_4
	+ \frac{1}{2} \int_{M} F \wedge \star F	- \int_{\Sigma_{\ast}} C \wedge \star F \, .
\end{equation}
Holding the metric fixed, the variation of $C_3$ in the equation above results in,

\begin{align*}
	\delta & \left( \frac{1}{2} \int F \wedge \star F \right) - \delta \int_{\Sigma_{\ast}} C \wedge \star F = \\
	& \int_{M} d(\delta C \wedge \star F) - \int_{M} \delta C \wedge d \star F - \int_{\Sigma_{\ast}} \delta C \wedge \star F - \int_{\Sigma_{\ast}} C \wedge \delta(\star F) 
\end{align*}
and the expression above simplifies to,

\begin{equation}
	\delta \left( \frac{1}{2} \int F \wedge \star F \right) - \delta \int_{\Sigma_{\ast}} C \wedge \star F = + \int_{M} \delta C \wedge d \star F \, .
	\label{HH5}
\end{equation}
This again gives the on shell equation of motion as,

\begin{equation}
	d \star F \, = \, 0 \quad \Longleftrightarrow \quad	\nabla_{\mu} F^{\mu\nu\rho\sigma} \, = \, 0 \quad \Longleftrightarrow \quad
	F_{\mu\nu\rho\sigma} \, = \, q \, \epsilon_{\mu\nu\rho\sigma} \, .
\end{equation}
However, note that the addition of the Legendre term actually flips the sign of the (on-shell) four form piece in equation \eqref{HH5},

\begin{equation}
	\frac{1}{2} \int F \wedge \star F - \int_{\Sigma_{\ast}} C \wedge \star F = \frac{1}{2} \int_{M} d(C \wedge \star F) + \frac{1}{2} \int_{M} C \wedge d \star F - \int_{\Sigma_{\ast}} C \wedge \star F 
\end{equation}
and therefore,

\begin{equation}
	\frac{1}{2} \int F \wedge \star F - \int_{\Sigma_{\ast}} C \wedge \star F =	\frac{1}{2} \int_{\Sigma_{\ast}} C \wedge \star F - \int_{\Sigma_{\ast}} C \wedge \star F = - \, \frac{1}{2} \int_{\Sigma_{\ast}} C \wedge \star F \, .
\end{equation}
Since $\star F \, = \, q$ is a 0-form on $\Sigma_{\ast}$, we have,

\begin{equation}
	- \, \frac{1}{2} \int_{\Sigma_{\ast}} C \wedge \star F = - \, \frac{q}{2} \int_{\Sigma_{\ast}} C = - \, \frac{q}{2} \int_{M} dC = - \, \frac{q}{2} \int_{M} q \, \epsilon = - \, \frac{q^2}{2} \int_{M} \sqrt{g} \, d^4 x \, .
\end{equation}
Therefore, the on-shell action (C.12) becomes,

\begin{equation}
	S_E[g,\bar{C}] = - \int \sqrt{g} \, d^4 x \, \frac{1}{16\pi G} (R - 2\Lambda_0)	- \frac{q^2}{2} \int_{M} \sqrt{g} \, d^4 x 
\end{equation}
and the metric saddle results in,

\begin{equation}
	G_{\mu\nu} = - (\Lambda_0 - 4\pi G q^2) \, \bigg|_{\Lambda_{\rm eff}} \, g_{\mu\nu} \, .
	\label{Wu Saddle}
\end{equation}
Therefore, we see that using Wu’s boundary prescription and the associated Legendre term, integrating out $C$ first and then varying $g$ results in the correct classical - saddle - solution obtained by Duff. Moreover, it can be shown that varying $g$ first and then imposing $F \, = \, q \, \epsilon$ also results in the same equation as above. Therefore, the two procedures now commute and Hawking’s conclusion for small $\Lambda_{\rm eff}$ holds with Duff’s condition on $\Lambda_0$ (i.e. $\Lambda_0 > 0$).

\end{document}